\documentclass[fleqn,usenatbib]{mnras}

\usepackage{newtxtext,newtxmath}

\usepackage[T1]{fontenc}
\usepackage{ae,aecompl}
\usepackage{threeparttable}

\usepackage{graphicx}	
\usepackage{amsmath}	
\usepackage[dvipsnames]{xcolor}

\defcitealias{Frohmaier17}{F17}

\newcommand{\flbox}{\ensuremath{F_\mathrm{box}}}

\title[The rates of massive stellar explosions from PTF]{From core collapse to superluminous: The rates of massive stellar explosions from the Palomar Transient Factory}

\author[C. Frohmaier et al.]{C. Frohmaier,$^{1}$\thanks{E-mail: chris.frohmaier@port.ac.uk}
C.~R.~Angus,$^{2,3}$ 
M.  Vincenzi,$^{1,4}$  
M. Sullivan,$^{2}$
M. Smith,$^{2,5}$
\newauthor
P.E. Nugent,$^{6,7}$
S. B. Cenko,$^{8,9}$
A. Gal-Yam,$^{10}$
S. R. Kulkarni,$^{11}$
N. M. Law$^{12}$
\newauthor
and R. M. Quimby$^{13,14}$
\\
$^{1}$Institute of Cosmology and Gravitation, University of Portsmouth, Portsmouth, PO1 3FX, UK\\
$^{2}$School of Physics and Astronomy, University of Southampton, Highfield, Southampton, SO17 1BJ, UK\\
$^{3}$DARK, Niels Bohr Institute, University of Copenhagen, Lyngbyvej 2, 2100 Copenhagen, Denmark\\
$^{4}$DISCnet Centre for Doctoral Training, University of Portsmouth, Portsmouth, PO1 3FX, UK\\
$^{5}$Université de Lyon, F-69622, Lyon, France; Université de Lyon 1, Villeurbanne; CNRS/IN2P3, Institut de Physique des Deux Infinis, Lyon\\
$^{6}$Department of Astronomy, University of California, Berkeley, CA, 94720-3411, USA\\
$^{7}$Lawrence Berkeley National Laboratory, Berkeley, CA, 94720, USA\\
$^{8}$Astrophysics Science Division, NASA Goddard Space Flight Center, Mail Code 661, Greenbelt, MD 20771, USA\\
$^{9}$Joint Space-Science Institute, University of Maryland, College Park, MD 20742, USA\\
$^{10}$Department of Particle Physics and Astrophysics, Weizmann Institute of Science, Rehovot 76100, Israel\\
$^{11}$Cahill Centre for Astrophysics, California Institute of Technology, 1200 East California Boulevard, Pasadena, CA 91125, USA\\
$^{12}$Department of Physics and Astronomy, University of North Carolina, Chapel Hill, NC 27599, USA\\
$^{13}$Department of Astronomy/Mount Laguna Observatory, San Diego State University, 5500 Campanile Drive, San Diego, CA 92812-1221, USA\\
$^{14}$Kavli Institute for the Physics and Mathematics of the Universe (WPI), The University of Tokyo Institutes for Advanced Study,\\ The University of Tokyo, Kashiwa, Chiba 277-8583, Japan
}

\date{Accepted XXX. Received YYY; in original form ZZZ}

\pubyear{2020}

\begin{document}
\label{firstpage}
\pagerange{\pageref{firstpage}--\pageref{lastpage}}
\maketitle

\begin{abstract}
We present measurements of the local core collapse supernova (SN) rate using SN discoveries from the Palomar Transient Factory (PTF). We use a Monte Carlo simulation of hundreds of millions of SN light curve realizations coupled with the detailed PTF survey detection efficiencies to forward-model the SN rates in PTF. Using a sample of 86 core collapse SNe, including 26 stripped-envelope SNe (SESNe), we show that the overall core collapse SN volumetric rate is $r^\mathrm{CC}_v=9.10_{-1.27}^{+1.56}\times10^{-5}\,\text{SNe yr}^{-1}\,\text{Mpc}^{-3}\, h_{70}^{3}$ at $ \langle z \rangle = 0.028$, and the SESN volumetric rate is $r^\mathrm{SE}_v=2.41_{-0.64}^{+0.81}\times10^{-5}\, \text{SNe yr}^{-1}\,\text{Mpc}^{-3}\, h_{70}^{3}$. We further measure a volumetric rate for hydrogen-free superluminous SNe (SLSNe-I) using 8 events at $z{\le}0.2$ of $r^\mathrm{SLSN-I}_v=35_{-13}^{+25}\, \text{SNe yr}^{-1}\text{Gpc}^{-3}\, h_{70}^{3}$, which represents the most precise SLSN-I rate measurement to date. Using a simple cosmic star-formation history to adjust these volumetric rate measurements to the same redshift, we measure a local ratio of SLSN-I to SESN of $\sim1/810^{+1500}_{-94}$, and of SLSN-I to all CCSN types of $\sim 1/3500^{+2800}_{-720}$. However, using host galaxy stellar mass as a proxy for metallicity, we also show that this ratio is strongly metallicity dependent: in low-mass ($\mathrm{log} M_{*} < 9.5 \mathrm{M}_\odot$) galaxies, which are the only environments that host SLSN-I in our sample, we measure a SLSN-I to SESN fraction of $1/300^{+380}_{-170}$ and  $1/1700^{+1800}_{-720}$ for all CCSN. We further investigate the SN rates a function of host galaxy stellar mass, and show that the specific rates of all core collapse SNe decrease with increasing stellar mass.

\end{abstract}



\section{Introduction}

The discovery space of optical transient surveys has been significantly expanded by recent wide field, high cadence, untargeted transient surveys. This has allowed the transient luminosity--duration domain to be populated at an unprecedented rate, uncovering an increasingly diverse variable universe, with new classes of events ranging from the fast-and-faint to the enduring-and-bright. This has also highlighted the diversity of supernova (SN) classes.

This is particularly  pronounced amongst core collapse SNe (CCSNe), the explosion of massive stars \citep[ $ \sim>8M_{\odot}$; e.g.,][]{Heger2003, Smartt2009}. Observationally, this group can be dichotomised based on the presence of hydrogen in their spectra, with \lq hydrogen-rich events\rq\ displaying clear signatures at all epochs of their evolution. The hydrogen-poor \lq stripped envelope\rq\ SNe (SESNe), on the other hand, predominantly lack hydrogen in their spectra (with the exception of SNe-IIb, which display some hydrogen at early times before evolving to show prominent helium emission later), and may or may not display signatures of helium (subclasses Ib and Ic respectively). These SESN subclasses are  heterogeneous \citep[e.g.][]{Lyman2014, Prentice2019, Shivvers2019}, although they can be broadly grouped based upon the relative abundances of H/He in their spectra \citep[e.g.][]{Modjaz2015,Prentice2017}. They are thought to arise due to progenitors which have been subject to differing levels of stripping prior to explosion. However, the exact progenitor systems of SESNe, and the way in which these outer layers of hydrogen-rich material are removed from the star, are currently unknown. Direct searches have identified a handful of SN IIb progenitors \citep[e.g.][]{Crockett2008,Smartt2009,Arcavi2011,Maund2011,Ryder2018,VanDyk2014,Kilpatrick2017,Tartaglia2017}, but for SN Ib/c, where the Wolf-Rayet (WR) progenitors are difficult to detect in optical imaging \citep{Smartt2009,Eldridge2012}, only one SN Ib \citep{2013ApJ...775L...7C} progenitor detection and one SN Ic detection \citep{2018ApJ...860...90V} in pre-explosion imaging.

Recently, an additional subclass of SESN events has emerged: hydrogen-poor superluminous supernovae (SLSNe-I). This class exhibits typically bright peak luminosities \citep[M$_g<-19.8$\,mag;][]{2019ARA&A..57..305G} and long-lived optical light curves, requiring extreme energies behind their production (events frequently radiate in excess of $\sim10^{51}\,{\rm ergs\,s^{-1}}$). Here too, there exists significant uncertainty surrounding their progenitor systems, as their energy requirements surpass the typical energies produced under standard core-collapse models. Proposed models for their production range from interaction with hydrogen-free circumstellar material \citep{Chevalier2011,Chatzopoulos2013,Sorokina2016}, to pair instability explosions \citep{Gal-Yam2009,Yan2015}, to energy injection from a central compact object \citep{Kasen2010,Woosley2010,Inserra2013}. The latter of these theories, implemented through the spin down of a newly-formed magnetar, is capable of broadly replicating the light curve evolution and spectroscopic features of a large number of SLSNe-I reasonably well \citep{Dessart2012,Inserra2013,Nicholl2013,Mazzali2016,Nicholl2017B}.

In more recent years, homogeneously selected samples of SLSNe have begun to highlight significant diversity in both their photometric and spectroscopic properties \citep{DeCia2018,Lunnan2018,Quimby2018,Angus2019}, demonstrating a much broader range of luminosities and evolutionary timescales than previously attributed to this spectroscopic class. Indeed, such samples have revealed fainter SLSNe, highlighting a potentially undersampled population of less luminous events in previous generation SN surveys. The luminosities and energies associated with these less-luminous SLSNe draw nearer to the brighter end of the \lq normal\rq\ CCSN regime. The spectral similarity of these events to classical SNe Ic  \citep{Pastorello2010} at late times \citep{Inserra2013} and during the nebular phase \citep{Nicholl2016B, Jerkstrand2017, 2019ApJ...871..102N}, has led to suggestions of a connection between the two spectroscopic subclasses \citep{2017ApJ...845...85L,DeCia2018,Quimby2018}.

Whether a physical connection exists between all the spectroscopic subclasses of SESNe in uncertain. A magnetar-powered scenario is able to provide a broad range of power outputs depending upon the initial spin period and magnetic-field strength of the magnetar \citep{Woosley2010,Kasen2010,Nicholl2017B} and as such, could provide a unified model to encompass all SESN explosions. Although there is little evidence, or consensus, that a magnetar is the dominant energy source over the decay of ${^{56}}\mathrm{Ni}$ in SESNe

One way in which we can begin to probe the underlying progenitor systems of SESNe is through an analysis of volumetric rates. For instance, for lone WR models of \lq normal\rq\ (Ib/c) SESN production, the relative rates of SN Ib/c to SN II should reflect the relative population fractions of WR stars at a given redshift. Conversely, for binary driven scenarios, the relative rate should reflect the population fractions of interacting massive stars within binaries.

Volumetric rates may also reflect any environmental dependencies in SESN progenitor production. An evolving rate with redshift may reflect an evolution of host galaxy properties (such as metallicity or star formation rate) with cosmic time, and thus provide clues as to the dependencies of their progenitors upon their host environment. Conversely, deviations from any expected environmental evolution at low/high redshift may reflect additional factors in progenitor production, which may have previously been overlooked. Understanding how the volumetric rate of events such as SLSNe-I evolve with redshift will not only improve searches for them in both current and upcoming surveys such as those conducted by \textit{Euclid} and the Vera Rubin Observatory \citep{Inserra2017Euclid,Villar2018}, but can also help provide a better understanding of their likely progenitor systems.

In this paper, we present the measurements of the rate of all CCSNe, stripped-envelope SNe sub-types and SLSNe-I from the Palomar Transient Factory \citep[PTF;][]{PTF_REF}. PTF was an automated optical sky transient survey operating at the Samuel Oschin 48 inch telescope (P48) at the Palomar Observatory, searching over 8000 deg$^{2}$ of the optical sky over its initial phase between 2009--2012, reaching typical depths of $M_{R}\sim20.5$. This wide-field survey presents the opportunity to constrain the low-redshift rate of core collapse events of all luminosities, from which we may compare to models of their production and any environmental dependencies in the local universe. 

In Section~\ref{sec:ratesPTF}, we begin with a description of the general method for calculating volumetric transient rates in the PTF survey. We also describe our method of constructing template light curves and justify our assumptions for the model parameters. In Section~\ref{sec:sample} we present our supernova sample for each of the transient sub-types observed by PTF. This is followed by a detailed explanation of the marriage between simulation and observation to calculate our rates of CCSNe and SLSNe in Section~\ref{sec:rates}. The CCSN-to-SLSN relative-rate implications are reviewed in Section~\ref{sec:relative}. Throughout, where relevant, we assume a flat $\Lambda$CDM Universe with a matter density $\Omega_{M}=0.3$ and a Hubble constant of $H_{0}= 70$km\,s$^{-1}$\,Mpc$^{-1}$, and we work in the AB photometric system.

\section{Rate calculation framework}
\label{sec:ratesPTF}

We begin by outlining the framework that we use for calculating the PTF SN rates.

\subsection{PTF survey simulations}

This work makes extensive use of the analysis of \citet[][hereafter F17]{Frohmaier17}, which produced the detection efficiencies of point sources in images processed by the real-time difference imaging pipeline of PTF data. This study inserted around 7 million artificial point sources into PTF images to test the performance of the PTF transient detection pipeline. They found that a combination of the observing conditions (seeing/image quality, sky brightness, and limiting magnitude), the transient's host galaxy surface brightness, and the transient's magnitude provided a good description of the survey's performance in detecting transients. From this, a multidimensional efficiency grid was created that described the probability that a point source (in an arbitrary environment) would have been detected in PTF difference imaging at any point in the survey.

We use these efficiencies in large Monte Carlo simulations to forward-model the PTF survey, with the input rate in our simulations adjusted until it matches the observed SN detection rate in the data \citep[e.g.][]{Prajs2016,Frohmaier18}. We use recent CCSN templates and luminosity functions \citep{Vincenzi2019} and SLSN template modelling \citep{Inserra2018,Angus2019} to simulate the artificial SLSNe. Key to the simulation is that real and simulated objects are treated identically, choosing the same sky areas (search volumes) and mimicking exactly the PTF survey strategy.

We define a sky area, $\Theta$, and an appropriate observing duration over which PTF would have been sensitive to our SNe of interest.
Since we are calculating volumetric rates, we define a (SN class dependent) maximum redshift out to which PTF could observe our SNe of interest. This is set by the maximum brightness of a typical object in the population, and by the light-curve evolution to ensure a sufficient number of observations for us to be confident of discovery.

Our methodology is as follows. We first draw a random value for the assumed intrinsic volumetric rate ($r_\mathrm{intrinsic}$), and use this to generate the number of events, $N_\mathrm{input}$, that would have occurred within a given volume ($V_\mathrm{obs}$) and observing duration ($T_\mathrm{obs}$) drawn from the Poisson distribution
\begin{equation}
    \label{eqn:poiSim}
    P(N_\mathrm{input}; \lambda)=\frac{\lambda^{N_\mathrm{input}} e^{-\lambda}}{N_\mathrm{input}!}
\end{equation}
where $\lambda=r_\mathrm{intrinsic}T_\mathrm{obs}V_\mathrm{obs}$, and
\begin{equation}
V_\mathrm{obs}=\frac{\Theta}{41253}\frac{4\pi}{3}\left[\frac{c}{H_0} \int_{0}^{z} \frac{dz'}{\sqrt{\Omega_M(1+z')^3+\Omega_\Lambda )}}\right ]^3\textrm{Mpc}^3,
\label{eqn:volume}
\end{equation}
where $c$ is the speed of light. Each object in the $N_\mathrm{input}$ sample is  assigned an explosion date, sky position, and a redshift such that they occur randomly and uniformly in the simulated volume-time space. The light curve properties for the $N_\mathrm{input}$ templates are drawn from a distribution of intrinsic population parameters described in Sections \ref{sec:simCCSN} and \ref{sec:simSLSN}. Milky Way extinction is applied to each template based on its location \citep{Schlegel1998} following \citet{Cardelli1989}. We include the detection efficiency effects of the host-galaxy surface brightness through the \flbox\ metric \citepalias[extensively detailed in][]{Frohmaier17}, measuring \flbox\ in pre-explosion imaging separately for all SNe sub-classes in our sample. This produces a set of host environments free from contaminant SN flux and that is representative of the observed host distribution. Each template is randomly assigned an \flbox\ from the population distribution and allows us to simulate, for example, faint supernovae in bright galaxies within the \citetalias{Frohmaier17} framework.

We then generate mock observations of our artificial SNe and use the observational efficiencies of \citetalias{Frohmaier17} to determine whether an object would have been detected by the PTF transient detection pipeline. Our simulation executes the observing strategy of PTF, and each time an artificial SN falls within an observation we use the SN template to calculate the brightness and, with the night's observing conditions, make a probabilistic statement as to whether that object would have been detected \citepalias[an example for type Ia SNe is presented in][]{Frohmaier17}. 

We require that our SN must have been \lq detected\rq\ on at least four nights over its duration, at which point we are confident a real object matching these requirements would have obtained a spectroscopic classification. The total number of objects passing this criterion, $N_\mathrm{output}$, represents the output for a single realization of our rate simulation. Repeating this process many times then builds up a description of the relationship between the intrinsic SN rate input into the simulation, and the observed number of SNe in the final sample. We then compare the $N_\mathrm{output}$ SNe to the observed populations of real SNe, linking back to the original distributions of $r_\mathrm{intrinsic}$ and, hence, the true SN rate. This method is dependent on realistic SN template light curves to produce reliable rate estimates, which we describe in the following section.

\subsection{Supernova Template Light Curves}
\label{sec:templates}
\subsubsection{Core-Collapse Supernovae}
\label{sec:simCCSN}

Our CCSNe are simulated using a library of 67 templates presented in \citet{Vincenzi2019}. This library has been built combining spectroscopy and multi-band photometry from well-observed CCSNe, and the templates span 6 classes of CCSN events (SN II, SN IIb, SN IIn, SN Ib, SN Ic and SN Ic-BL). Each individual template extends from 1700\AA\ to 11000\AA. Every CCSN sub-class also has an associated luminosity function, such that every template is adjusted so that the peak brightness of the simulated events follows this function \citep[see][for further details]{Vincenzi2019}. Furthermore, we adopt the host-galaxy extincted version of these models and assume the reddening in the templates is representative of the reddening in the data.

We additionally separate out the templates associated with SESNe (i.e. SNe Ib, SNe Ic, SNe Ic-BL and SNe IIb) to calculate a SESN only rate. We do not differentiate between the sub-classes of these normal luminosity SESNe, as divisions between these sub-classes are poorly defined \citep{Shivvers2019}, and mis-classification  is a potential contaminant to any subsequent rate calculation.

\subsubsection{Superluminous Supernovae}
\label{sec:simSLSN}

Due to the scarcity of spectrophotometric data for SLSNe-I we create our template light curves using a parametric model capable of replicating the general form of a SLSN-I event. This neglects the small-scale behaviours often present within SLSN light curves, which may be indicative of multiple energy sources \citep[e.g.][]{Inserra2018}, as these features do not significantly affect the detectability of simulated SLSNe-I events.

We adopt the magnetar model of \citet{Inserra2013}, including a time dependant trapping coefficient introduced by \cite{Wange2015} to account for late-time behaviour. Models of magnetar spin-down have been shown to replicate the broad photometric properties of a large number of SLSN-I light curves \citep{Inserra2013,Nicholl2013,Nicholl2017B,Dessart2019}, and whilst there remains  debate over whether a magnetar is the sole power source \citep[e.g.][]{Inserra2018,Angus2019}, it provides a framework with which a \lq typical\rq\ SLSN light curve can be generated. The luminosity is given by

\begin{multline}
L_{\mathrm{SN}}\left ( t \right )= \left(1- \exp{\left(\frac{9\kappa M_{ej}^{2}}{40\pi E_{k}}t^{-2} \right)}\right)\times e^{-(t/\tau_{\mathrm{m}})^{2}} \\ 2\int_{0}^{t/\tau_{\mathrm{m}}}
\left(\int 4.9\times10^{46} \left(\frac{B}{10^{14}~\mathrm{G}} \right)^{2}\left(\frac{P}{\mathrm{ms}} \right)^{-4} \frac{1}{(1 + t/\tau_{\mathrm{p}})^{2}}\right ) \\ e^{(t^{\prime}/\tau_{\mathrm{m}})^{2}}\frac{dt^{\prime}}{\tau_{\mathrm{m}}} \mathrm{~ erg~ s^{-1}},
\label{equation_magnetar}
\end{multline}
where $B$ and $P$ are the magnetic field strength and period of the magnetar, respectively, $\tau_{\mathrm{p}}$ is the spin down timescale of the magnetar, and $\tau_{\mathrm{m}}$ is the diffusion timescale, which under the assumption of uniform ejecta density, can be expressed in terms of the mass, $M_{\mathrm{ej}}$, opacity, $\kappa$, and kinetic energy, $E_{\mathrm{k}}$ of the ejecta. We assume an explosion energy of 10$^{51}$ erg and a hydrogen-free ejecta with an opacity of $\kappa = 0.1~\mathrm{cm}^{2} \mathrm{g}^{-1}$ \citep{Inserra2013}. We explored an opacity of $\kappa = 0.01~\mathrm{cm}^{2} \mathrm{g}^{-1}$, which is supported by late-time SLSNe observations \citep{2018ApJ...866L..24N}. However the inclusion of this parameter value has little affect upon the results presented here, as the influence of this lower opacity does not become predominant until extremely late times  ($\gtrapprox250$ days post maximum brightness).

We generate SLSN-I light curves by varying values $B$, $P$ and $\tau_{\mathrm{m}}$. To explore the parameter-space, we fit this model to the light curves of a well-observed sample of SLSNe-I with $z<0.3$, close to the $z<0.2$ redshift range of our rates measurement. We place the following data requirements on this sample to guarantee adequate sampling across both the duration of the light curve and the underlying spectral energy distribution (SED) for each event:
\begin{enumerate}
    \item The light curves must have data in $>$3  photometric bands to constrain the underlying SED;
    \item The light curve must be well-sampled in these bands ($>$6 epochs of data in each band).
\end{enumerate}
We use Gaussian Processes to interpolate the literature-sample SLSN-I light curves onto a 1 day cadence grid, and fit the resulting interpolated light curves with the phase dependent UV-absorbed black body model of \citet{Inserra2017Euclid} and \citet{Angus2019}. At each epoch the SED is integrated to determine the bolometric luminosity of the SN, and the resulting bolometric light curves fit with the magnetar model outlined above. The literature sample of 20 objects are presented in Table~\ref{tab:slsnproperties} alongside the best-fit $B$, $P$ and $\tau_{\mathrm{m}}$ values. Data were obtained from The Open Supernova Catalogue \citep{2017ApJ...835...64G} using all SLSNe-I that met our sampling and redshift requirements.

\begin{table}
	\caption{The best-fit magnetar model parameters for a literature sample of low-redshift SLSNe in increasing redshift order.}
	\centering
	\begin{tabular}{l l l l l} 
		\hline
		SN name 	&  $z$  	&  $P_{\mathrm{spin}}$	 &$B$& $\tau_{\mathrm{M}}$	 \\
		  		&       & (ms) 				& ($10^{14}$G)& (days)			 \\
		\hline
        SN2018bsz & 0.026665	& 3.70	& 18.46	& 18.47 \\
        SN2017egm & 0.030721	& 15.41	& 5.33	& 11.58 \\	
        SN2018hti & 0.063		& 14.31	& 7.69	& 15.40 \\	
        SN2016eay & 0.1013		& 3.67	& 2.45	& 12.54 \\
        PTF12dam    &0.1073&     9.80 &	2.40 &	14.45 \\	
        SN2015bn    &0.1136 &     3.57  &  2.90 & 15.52 \\ 
        SN2017dwh & 0.13		& 12.95	& 18.24	& 13.52 \\		
        SN2018avk & 0.132		& 3.76	& 2.21	& 11.24 \\	
        SN2011ke    &0.1428&     6.50 &	10.31 &	12.60 \\
        SN2012il    &0.175&     13.84 &  12.44 & 14.72 \\
        PTF12gty  & 0.1768		& 7.34	& 1.94	& 12.12 \\	
        PTF09as     &0.1867 &     19.99 &	30.00 &	10.26 \\
        SN2011kg    &0.1924&     9.71  &  8.40 &14.25 \\
        SN2016ard & 0.2025		& 7.91	& 11.13	& 13.49 \\
        PTF10aagc   & 0.2067&     14.18 &	25.18 &11.12 \\ 
        SN2010gx    &0.2297&     7.73  &  6.68 & 13.59  \\
        LSQ14mo     &0.253&     9.26  &  8.50 &	13.86 \\
        PTF09cnd    &0.2584&     4.53  &  2.43 &14.62 \\
        SN2013dg  & 0.265		& 10.63	& 6.861	& 16.67 \\	
        SN2018bym & 0.267		& 7.37	& 5.75	& 14.12 \\
		\hline
		\label{tab:slsnproperties}
	\end{tabular}
 \end{table}

\begin{figure}
	\includegraphics[width=\columnwidth]{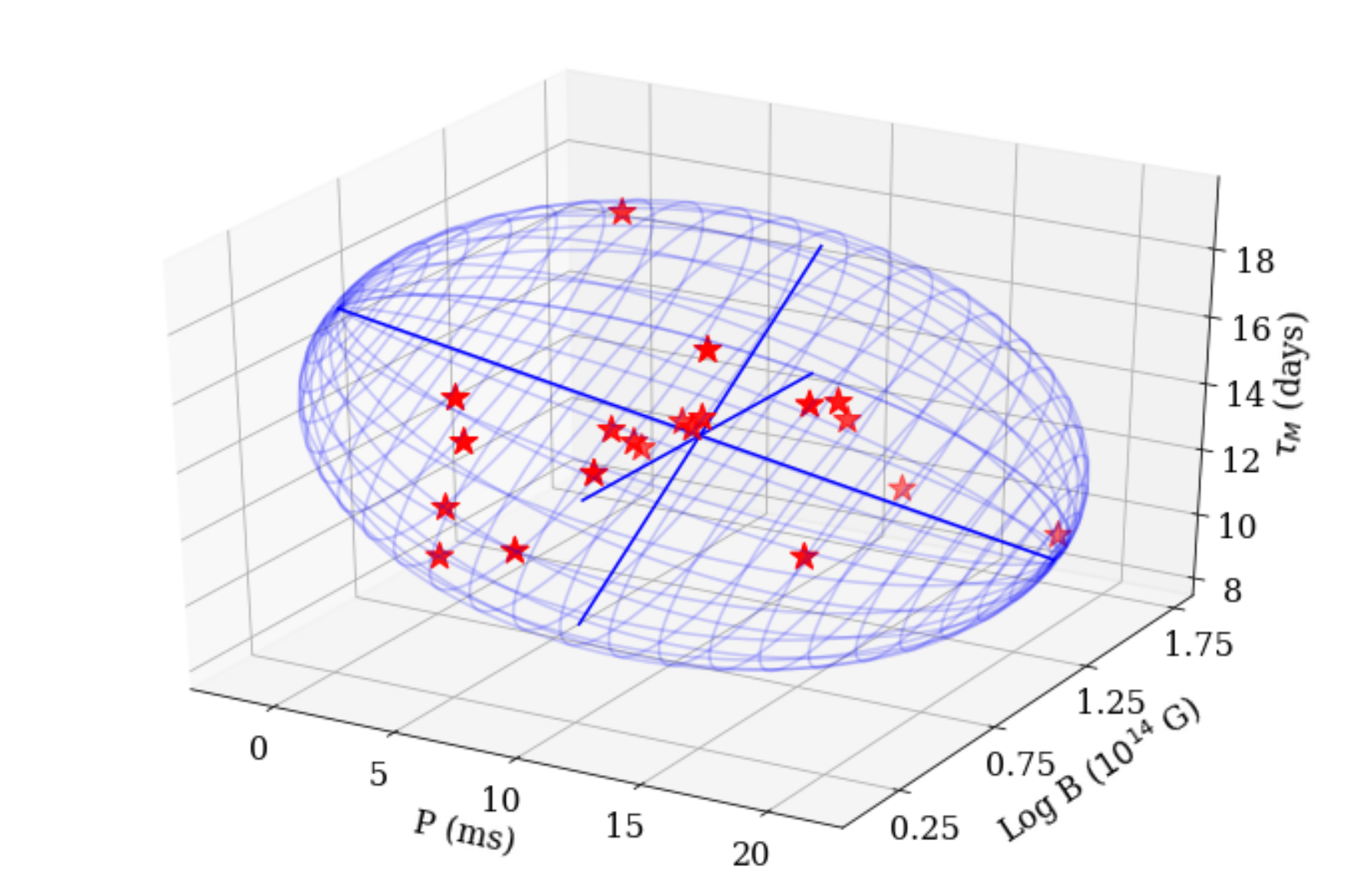}
    \caption{The $\tau_{M}$-B$_{14}$-P$_{ms}$ magnetar parameter space determined for the low-$z$ template SLSNe sample (red stars) and the smallest volume enclosing them (blue wire-frame).}
    \label{fig:param_space}
\end{figure}

We generate template light curves based models whose properties are contained within a parameter space enclosing the fitted properties of our sample. We follow \citet{Prajs2016} and use a Khachiyan algorithm \citep{Aspvall1980,Khachiyan1980} to determine the smallest volume which contains all of our fitted magnetars. We then create templates from fixed combinations of magnetar properties contained in coordinates within this ellipsoidal parameter space. Given the small number of objects used to define this model parameter space, we perform jackknife resampling to ensure the parameter space is not skewed by outlying points, and find our ellipsoid fit to be optimal for the sample. We visualise this final parameter space in Fig.~\ref{fig:param_space}. 

\begin{figure}
	\includegraphics[width=\columnwidth]{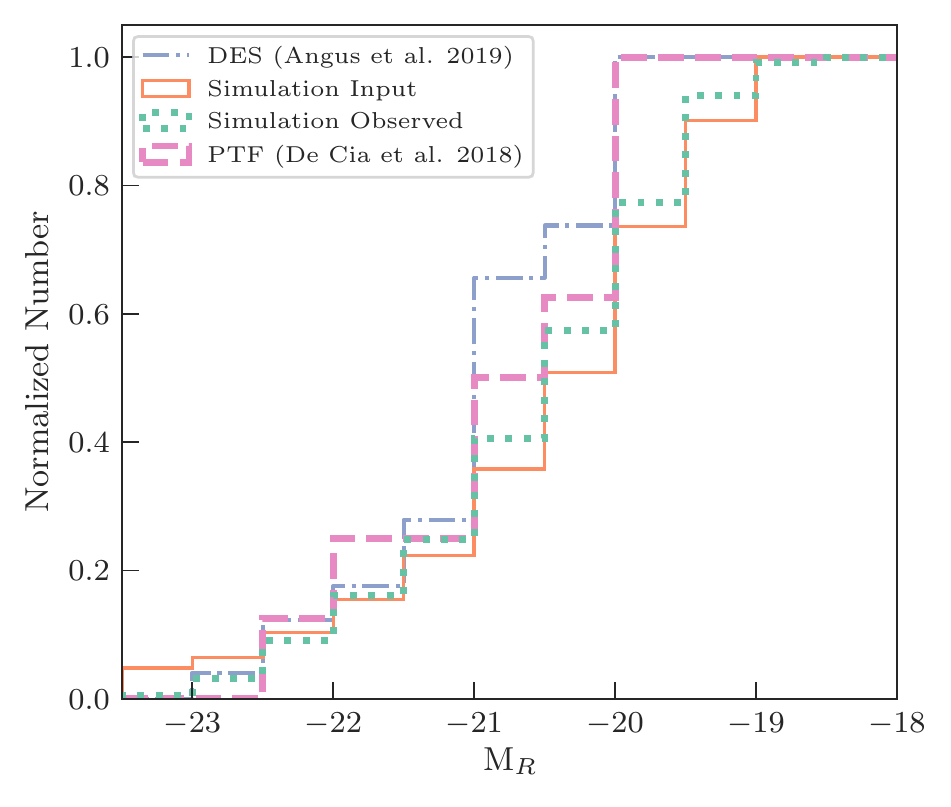}
    \caption{The cumulative distribution for the luminosity function of the \citet[][]{Angus2019} SLSN-I sample (dot-dashed line) and the PTF sample \citep[dashed line;][]{DeCia2018} from which we draw our rates sample. The solid line represents the luminosity function of our magnetar model derived from a literature population of well-observed SLSN-I (Table~\ref{tab:slsn_properties}). The dotted-line shows the luminosity function of the simulated set of SLSNe-I that have passed our observing requirements described in Section~\ref{sec:rates}.}
    \label{fig:lit_LF}
\end{figure}

We finally create a library of template PTF $R$-band light curves, to phases of +100 days in the rest-frame, using the possible magnetars contained within this defined parameter space. The observed luminosity-function from our simulations in Figure~\ref{fig:lit_LF} is broadly representative of literature examples \citep[e.g.][]{Angus2019}; although all homogeneously selected SLSN-I samples suffer from low number statistics and Malmquist bias. We assume a UV absorbed black body as the underlying SED and generate observer-frame light curves over all redshifts within the range $0.001<z<0.2$, and a redshift bin size of 0.001.

\section{The PTF supernova sample}
\label{sec:sample}

This section details the construction of our CCSN and SLSN samples. The largest uncertainty in our rate calculation is the spectroscopic completeness: the fraction of SNe detected by PTF that have a spectroscopic type and redshift. To mitigate this, our simulation parameters are designed to include times when PTF achieved a regular cadence and high spectroscopic classification efficiency, thus minimising potential contaminants in the final sample. 

We first describe our light curve coverage cuts and discuss the details of our search for both spectroscopically confirmed and photometrically identified candidates. All cuts are applied to both our real and simulated SNe.

\subsection{Light-curve coverage cuts}
\label{sec:coverage_cuts}

We adopt some of the light-curve coverage cuts made by \citet{Frohmaier19} in their calculation of the PTF SN Ia rate. Our coverage cuts are:
\begin{enumerate}
    \item There must be at least four epochs on which the SN is detected. To be considered detected, each epoch must have an RB score $\ge$0.07 \citep{Bloom2012};
    \item These four epochs must all be separated by $\ge$12 h.
\end{enumerate}
These cuts were applied to the real-time PTF difference-imaging photometry\footnote{A publicly available replica database, containing sources and photometry, can be accessed from \url{https://irsa.ipac.caltech.edu/} }.

\begin{figure}
	\includegraphics[width=\columnwidth]{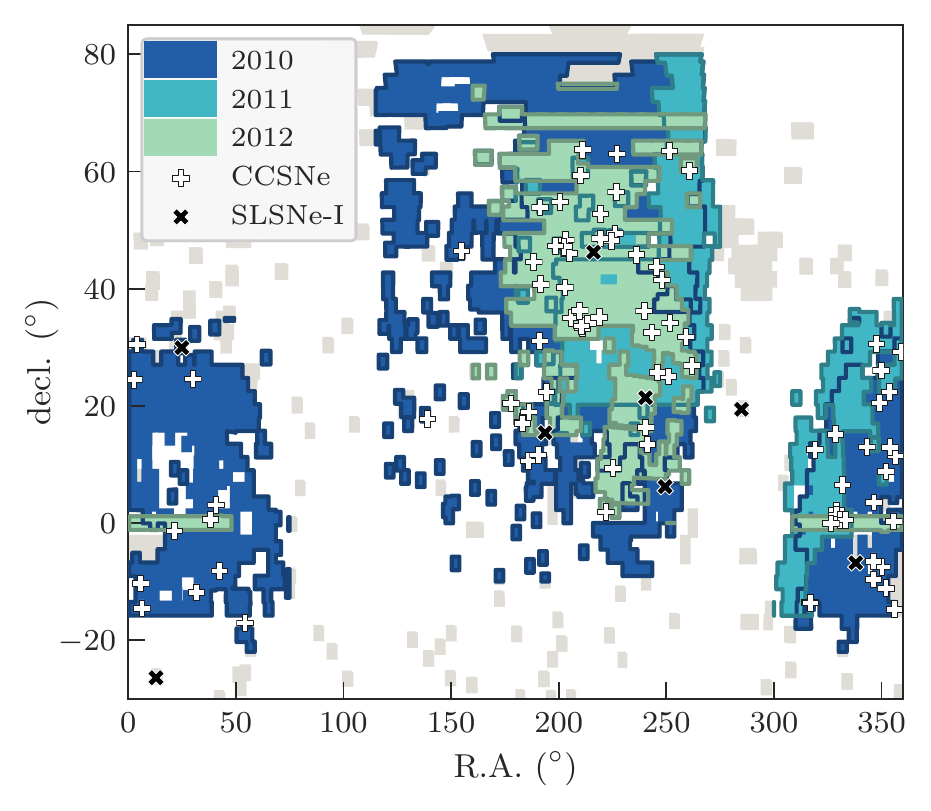}
    \caption{The PTF-observed footprint for the three years (2010--2012) of our simulations. The CCSN rate sub-survey areas are colour-coded for each year of the survey, while the SLSNe-I fields include the greyed-out tiles for all possible PTF observations.  Over-plotted is the spatial distribution of the PTF-discovered SLSNe-I and CCSNe. }
    \label{fig:ptfSky}
\end{figure}

\subsection{The core-collapse sample}
\label{sec:CCSample}

For our CCSN sample, we set a redshift limit of z${\le}$0.035, chosen to ensure that objects at the faint end of the CCSN luminosity function \citep{2011MNRAS.412.1441L} are detectable by PTF (m ${\lesssim}$20.5). We then defined the sky area and date range over which to perform a search for real CCSNe. 

The PTF search strategy evolved over the survey lifetime, which means that there is no single area that can be used over the three years of PTF operation. We instead establish 9 areas across 2010--2012 where PTF sustained a regular cadence so that we could model each as a distinct sub-survey. These areas were selected through a visual inspection of the PTF fields and were unbiased by the SN distribution on the sky. The total sky is visualized in Fig.~\ref{fig:ptfSky} and broken-down by the year of observation. The individual sub-survey parameters are listed in Table~\ref{tab:lc_sky_area_nights}. We then searched the PTF database to find SN candidates within these sub-surveys.

\subsubsection{Spectroscopically confirmed CCSN}
\label{sec:spec_CC}

The spectroscopic follow-up campaign for the PTF survey is extensively detailed in the literature \citep[e.g.,][]{Frohmaier19}, and we do not reproduce these details here. CCSNe in PTF were classified into  sub-types based on their spectral features: SNe II (strong H lines), SNe IIb (strong H and He lines), SNe IIn (narrow H), SNe Ib (no H, strong He lines), and SNe Ic (no H or He lines). Additionally, events were classified as SN Ic-BL based on their similarities to SN 1998bw and SN 2002ap. Our SNe II were not sub-divided further into light curve evolution driven categories (IIP/L groups), although we do distinguish SN IIb events as they are believed to originate from a stripped progenitor \citep[e.g.][]{2011ApJ...739...41C}. In total, 84 CCSNe (60 SNe II, 24 SESNe) were classified that also pass our light curve coverage cuts.

\subsubsection{Photometrically Identified CCSNe}
\label{sec:photCC}

We next search the PTF transient database of $\sim48000$ objects for candidate CCSNe that were not observed spectroscopically, following closely the method of \citet{Frohmaier19}. After applying our light curve coverage cuts, and removing those objects with a spectroscopic classification, there are 20 PTF objects that meet our criteria and have no spectroscopic information. These were cross-matched with the SDSS galaxy catalogues, and objects with $z_\mathrm{host}>0.035$ were removed. This left three objects: PTF12bvf, PTF12cxz, and PTF12gfx. PTF12cxz is coincident with a known SN-imposter candidate that was later identified in 2014 \citep{2014ATel.5737....1T}, and is thus likely a luminous blue variable eruption, and so excluded from the sample. This leaves two events -- PTF12bvf and PTF12gfx -- with host spectroscopic redshifts within our range. A visual inspection of the light curves and faint peak magnitudes (M$_\mathrm{R} > -17$) rule out a SN Ia classification. We consider these objects to be viable CCSN candidates and include them in this analysis. Our final sample of PTF spectroscopic and photometric CCSNe are listed in Table~\ref{tab:ccsn_properties}.

\begin{table*}
    \caption{The PTF CCSN sample used in our calculation. Our sample contains 86 CCSNe of which 24 are SESNe. The two photometrically-identified candidates do not have an assigned classification.}
    \label{tab:ccsn_properties}
    \begin{center}
    \begin{tabular}{lcccrl}
        \hline
    PTF Name & IAU Name & R.A. & decl. & Redshift & SN Type \\
             &          & (J2000) & (J2000) & & \\
        \hline
    PTF10acbu & - & 14:00:46.92 & +59:20:16.53 & 0.0099 & SN Ic \\
PTF10bau & 2010Z & 09:16:21.29 & +17:43:40.24 & 0.026 & SN IIP \\
PTF10bgl & - & 10:19:04.70 & +46:27:23.34 & 0.03 & SN IIP \\
PTF10bld & - & 12:12:16.38 & +17:14:45.52 & 0.027 & SN II \\
PTF10cxx & - & 14:47:27.78 & +01:55:03.79 & 0.034 & SN IIb \\
PTF10dvb & 2010aw & 17:16:12.25 & +31:47:36.01 & 0.023 & SN IIP \\
PTF10ehy & - & 13:39:29.47 & +46:00:53.75 & 0.006 & SN IIP \\
PTF10eqe & - & 14:04:34.18 & +63:43:26.15 & 0.031 & SN II \\
PTF10eqi & 2010av & 15:06:56.97 & +56:30:26.47 & 0.03 & SN Ib \\
PTF10eqz & - & 15:08:10.43 & +63:00:42.49 & 0.029 & SN II \\
PTF10feq & - & 12:12:21.40 & +16:58:45.17 & 0.028 & SN Ib \\
PTF10fgz & - & 14:37:44.53 & +52:43:32.70 & 0.012 & SN II \\
PTF10fjh & 2010bq & 16:46:55.36 & +34:09:34.71 & 0.032 & SN IIn \\
PTF10fmr & - & 12:33:24.36 & +44:32:16.04 & 0.02 & SN IIb \\
PTF10fqg & - & 12:41:49.86 & +11:35:28.11 & 0.028 & SN IIb \\
PTF10fqx & - & 16:46:01.27 & +63:28:59.52 & 0.017 & SN IIP \\
PTF10gki & 2010ck & 14:04:11.25 & +33:18:21.61 & 0.026 & SN II \\
PTF10gva & - & 12:23:55.40 & +10:34:50.62 & 0.028 & SN II \\
PTF10gxi & - & 12:44:33.68 & +31:05:05.35 & 0.029 & SN II \\
PTF10hlc & - & 15:44:18.60 & +45:48:14.97 & 0.035 & SN II \\
PTF10hny & - & 15:04:25.29 & +49:24:02.96 & 0.027 & SN IIP \\
PTF10hpa & 2010ea & 12:45:29.30 & +53:51:47.37 & 0.03 & SN IIP \\
PTF10hyq & - & 12:46:27.24 & +40:45:00.32 & 0.016 & SN IIP \\
PTF10jkx & - & 13:32:57.11 & +48:18:54.55 & 0.029 & SN II \\
PTF10kui & - & 13:31:38.54 & +40:11:33.96 & 0.021 & SN Ib \\
PTF10lnx & - & 16:04:39.18 & +13:23:43.68 & 0.034 & SN II \\
PTF10myz & - & 14:58:13.71 & +48:11:28.19 & 0.028 & SN IIP \\
PTF10npd & - & 23:20:12.62 & -07:28:50.42 & 0.01 & SN IIP \\
PTF10osr & - & 23:45:45.16 & +11:28:42.37 & 0.024 & SN IIP \\
PTF10qob & 2010ib & 01:25:47.51 & -01:22:30.13 & 0.019 & SN IIP \\
PTF10qwz & 2010hm & 23:35:18.61 & +12:55:31.81 & 0.02 & SN IIP \\
PTF10rad & - & 23:28:24.08 & -11:07:27.67 & 0.024 & SN II \\
PTF10raj & - & 02:43:24.13 & +03:06:17.58 & 0.023 & SN II \\
PTF10rin & - & 22:52:38.95 & +13:00:36.57 & 0.032 & SN IIP \\
PTF10rmn & - & 00:16:25.40 & +30:28:51.29 & 0.024 & SN IIP \\
PTF10svt & - & 03:37:17.83 & -17:05:41.23 & 0.031 & SN Ic \\
PTF10tff & - & 14:37:01.40 & +48:32:40.62 & 0.0075 & SN IIP \\
PTF10tpa & - & 23:04:40.96 & -09:38:27.01 & 0.033 & SN IIP \\
PTF10ttd & - & 02:06:52.17 & -11:49:57.04 & 0.015 & SN IIP \\
PTF10uiy & - & 00:23:16.42 & -10:18:54.68 & 0.023 & SN II \\
PTF10ujc & - & 23:34:31.92 & +22:21:03.24 & 0.032 & SN IIn \\
PTF10vdl & - & 23:05:48.88 & +03:31:25.54 & 0.016 & SN IIP \\
PTF10vnv & 2010ha & 02:00:24.81 & +24:34:44.16 & 0.015 & SN Ib \\
PTF10wcy & - & 00:25:19.50 & -14:33:37.50 & 0.024 & SN II \\
PTF10wmf & - & 21:56:15.70 & +02:10:13.78 & 0.029 & SN IIP \\
PTF10xgo & - & 21:55:57.38 & +01:19:14.11 & 0.034 & SN IIn \\
PTF10xjr & - & 02:49:19.49 & -08:10:30.21 & 0.03 & SN Ib \\
PTF10xjv & - & 00:10:29.13 & +24:28:01.05 & 0.028 & SN IIP \\
PTF10yna & - & 23:15:53.57 & +20:28:17.36 & 0.023 & SN II \\
PTF10yow & 2010iq & 21:54:23.30 & +15:09:20.65 & 0.025 & SN Ic \\
PTF10zcn & 2010is & 23:19:14.39 & +26:03:11.62 & 0.02 & SN Ic \\
PTF11bli & - & 14:02:16.18 & +33:39:41.46 & 0.03 & SN Ib/c \\
PTF11bov & 2011bm & 12:56:53.94 & +22:22:28.14 & 0.022 & SN Ic \\
PTF11cgx & - & 14:13:46.00 & +34:02:12.93 & 0.033 & SN II \\
PTF11dhf & - & 17:22:51.47 & +60:07:49.88 & 0.028 & SN IIb \\
PTF11ecp & - & 16:43:54.10 & +25:00:53.98 & 0.034 & SN II \\
PTF11eon & 2011dh & 13:30:05.08 & +47:10:11.22 & 0.0016 & SN IIb \\
PTF11ftc & - & 21:06:39.24 & -13:33:09.46 & 0.03 & SN II \\

        \hline

    \end{tabular}
    \end{center}
    \end{table*}
    
    \begin{table*}
    \contcaption{The PTF CCSN sample used in our calculation. Our sample contains 86 CCSNe of which 24 are SESNe. The two photometrically-identified candidates do not have an assigned classification.}
    \label{tab:cont_ccsn_properties}
    \begin{center}
     \begin{tabular}{lcccrl}
        \hline
    PTF Name & IAU Name & R.A. & decl. & Redshift & SN Type \\
             &          & (J2000) & (J2000) & & \\
        \hline
PTF11ftr & - & 23:55:35.79 & +29:12:19.94 & 0.018 & SN II \\
PTF11fuv & 2011kh & 22:03:15.00 & +00:34:24.90 & 0.03 & SN IIP \\
PTF11gls & 2011dw & 16:31:39.38 & +41:29:23.12 & 0.03 & SN IIP \\
PTF11htj & - & 21:16:03.50 & +12:31:20.95 & 0.017 & SN IIP \\
PTF11hyg & 2011ee & 23:27:57.34 & +08:46:38.02 & 0.03 & SN Ic \\
PTF11iil & - & 21:07:41.58 & -13:39:55.11 & 0.029 & SN II \\
PTF11klg & - & 22:07:09.92 & +06:29:08.72 & 0.027 & SN Ic \\
PTF11qax & - & 23:42:25.58 & +00:15:16.83 & 0.022 & SN IIP \\
PTF11qcj & - & 13:13:41.51 & +47:17:57.03 & 0.028 & SN Ibn \\
PTF12bro & 2012br & 12:24:17.05 & +18:55:27.96 & 0.023 & SN IIP \\
PTF12bvf & - & 11:51:01.66 & +20:23:59.64 & 0.022 & SN \\
PTF12cde & - & 13:58:58.39 & +36:14:26.43 & 0.013 & SN Ib/c \\
PTF12cgb & - & 16:21:44.43 & +43:44:23.15 & 0.026 & SN II \\
PTF12cli & - & 13:42:34.46 & +35:01:15.83 & 0.024 & SN II \\
PTF12cod & 2012cd & 13:22:35.29 & +54:48:47.11 & 0.012 & SN II \\
PTF12dcp & 2012bw & 16:12:56.12 & +32:30:43.21 & 0.031 & SN Ic \\
PTF12eaw & - & 14:35:17.71 & +35:07:04.14 & 0.029 & SN Ib \\
PTF12fip & - & 15:00:51.04 & +09:20:25.12 & 0.034 & SN II \\
PTF12gfx & - & 16:01:39.35 & +16:18:36.88 & 0.029 & SN \\
PTF12gnn & - & 15:58:49.28 & +36:10:10.95 & 0.031 & SN II \\
PTF12gnt & - & 17:27:47.30 & +26:51:22.11 & 0.029 & SN II \\
PTF12gzk & - & 22:12:41.53 & +00:30:43.07 & 0.014 & SN Ic \\
PTF12hap & - & 16:23:10.01 & +25:42:06.92 & 0.02 & SN II \\
PTF12hno & - & 23:04:55.63 & -06:32:11.67 & 0.019 & SN II \\
PTF12hvv & - & 21:45:46.45 & -00:03:25.12 & 0.029 & SN Ic \\
PTF12iif & - & 23:43:44.15 & -14:42:38.30 & 0.024 & SN IIP \\
PTF12izy & 2012iw & 02:32:39.67 & +00:37:00.12 & 0.021 & SN II \\
PTF12jfp & 2012fc & 23:10:12.80 & +30:34:43.11 & 0.02 & SN II \\

    \hline

    \end{tabular}
    \end{center}
    \end{table*}

\subsection{The SLSN sample}
\label{SLSN_sample}

The PTF SLSN-I sample was classified in \citet{Quimby2018} and the light curves presented in \citet{DeCia2018}. In total, 26 SLSNe-I were studied to form the largest low redshift, homogeneously selected, sample of SLSN-I to-date.

We draw our candidates from \citet[][table 1.]{DeCia2018}. Each light curve was checked against the coverage cuts in Section~\ref{sec:coverage_cuts}. This coverage cut was designed to maximise the spectroscopic completeness for the SN Ia rate \citep{Frohmaier19}. The follow-up of targets was decided by observers at a number of telescopes, which may be biased towards different science goals (e.g. SN Ia). Naturally, this results in a complex selection function ultimately dictated by the brightness of the target transients. Motivated by this, as with the CCSNe, we make a redshift cut to ensure a high sample completeness. We choose to be conservative and set this as $z\le0.2$, below the volume-weighted mean redshift of the whole PTF sample at $\langle \mathrm{z} \rangle = 0.33$. This redshift limit coincides with an apparent peak in the \citet{DeCia2018} redshift distribution and ensures the the majority of objects simulated from our own luminosity function in Section~\ref{sec:simSLSN} will be detectable by PTF. Applying this redshift cut to the \citet{DeCia2018} sample provides us with 8 objects, which we list in Table \ref{tab:slsn_properties}.

The inclusion of PTF10hgi in our sample may be somewhat contentious as it has been classified as both SLSN-I \citep{Inserra2013, DeCia2018} and SLSN-IIb \citep{Quimby2018,2019ARA&A..57..305G} in the literature. Given that our sample was initially defined following the \citet{DeCia2018} analysis, we continue our rate measurement with PTF10hgi included. If, however, this object truly does not belong to the SLSN-I class, our resulting rate measurement may be a small over-estimate.  

We also included PTF10vwg (SN2010hy) in our sample, initially identified by the Lick Observatory Supernova Search \citep[LOSS;][]{LOSSref} as a high-luminosity SN Ic (although the possibility of a SN Ia classification was difficult to rule out). As noted by \citet{Quimby2018}, this object was located in a PTF field which was not subject to prompt searching due to its low galactic latitude $(b \approx 7\degr)$. PTF10vwg was later identified within the PTF data following its initial announcement, with follow-up observations showing it to be consistent with a SLSN-I. As with all rate studies, if other such objects exist in the data but were never formally discovered, the resulting measurements would be an underestimate of the intrinsic rate.

\begin{table*}
    \caption{The spectroscopically confirmed PTF SLSN sample used for this rate measurement.}
    \begin{center}
    \begin{tabular}{lcccrl}
        \hline
    PTF Name &  IAU Name & R.A. & decl. & Redshift & SN Type \\
             &   & (J2000) & (J2000) & & \\
        \hline
    PTF10bfz & - & 12:54:41.27 & +15:24:16.99 & 0.17 & SLSN-I \\
    PTF10hgi & 2010md & 16:37:47.04 & +06:12:32.32 & 0.099 & SLSN-I \\
    PTF10vwg & 2010hy & 18:59:32.86 & +19:24:25.74 & 0.19 & SLSN-I \\
    PTF11hrq & - & 00:51:47.22 & -26:25:09.99 & 0.057 & SLSN-I \\
    PTF11rks & 2011kg & 01:39:45.51 & +29:55:27.01 & 0.19 & SLSN-I \\
    PTF12dam & - & 14:24:46.20 & +46:13:48.32 & 0.11 & SLSN-I \\
    PTF12gty & - & 16:01:15.23 & +21:23:17.42 & 0.18 & SLSN-I \\
    PTF12hni & - & 22:31:55.86 & -06:47:48.99 & 0.11 & SLSN-I \\
    \hline
		\label{tab:slsn_properties}
    \end{tabular}
    \end{center}
    \end{table*}

\section{The volumetric SN rates}
\label{sec:rates}

In this section, we detail the process of calculating the CCSN and SLSN-I rates from the PTF data. We follow the methodology principles laid out in Section~\ref{sec:ratesPTF} and describe the subtle differences in the treatment of data for each of our calculated rates. We also present the measurement uncertainties and justify how our method naturally incorporates the statistical and systematic uncertainties into a single probability density function. We begin by calculating the rate of CCSNe and follow this by extracting only the SESNe for a separate analysis. We then perform a near identical study on the SLSNe-I and conclude by comparing all sub-sets to investigate the relative fractions.

\subsection{Core-Collapse Supernova Rate}
\label{sec:CCSN_Rate}

To calculate the volumetric rate, we start with a large Monte Carlo simulation where many realizations of an intrinsic SN rate, $r_\mathrm{intrinsic}$, are drawn along with the fixed $T_\mathrm{obs}$ and $V_\mathrm{obs}$ for the sub-surveys. These values are used in Equation~\ref{eqn:poiSim} to generate an $N_\mathrm{input}$ and, following the methods in Section~\ref{sec:ratesPTF}, a subsequent value of $N_\mathrm{output}$ is determined. This process was repeated for 400 000 realizations of the intrinsic CCSN rate, $r_\mathrm{intrinsic}$, to produce a corresponding set of $N_\mathrm{output}$. At this stage, it is possible to connect any value of $N_\mathrm{output}$ to a distribution of initial $r_\mathrm{intrinsic}$ \citep[e.g.][]{Prajs2016} values, however, we must also include an uncertainty on our true SN sample size.

The observed number of SN events follows an intrinsic rate (for any given volume and time-span) that is governed by the statistics of the Poisson point-process. The probability of observing an integer $n$ events given we have $N_\mathrm{obs}=86$ observed events follows the Poisson distribution, identical in functional form to Equation~\ref{eqn:poiSim} 

\begin{equation}
    P(n; N_\mathrm{obs})=\frac{N_\mathrm{obs}^{n} e^{-N_\mathrm{obs}}}{n!}
    \label{eqn:poiObs}
\end{equation}

In principle, n extends over the range $0 \le n \le \infty$, however, in practise $P(n=N_\mathrm{obs})$ quickly becomes vanishingly small either side of the maximum probability for our sample size of $N_\mathrm{obs}=86$. We therefore compute the probabilities over a finite range of n where beyond this the probabilities have a negligible effect upon the final result.

\begin{figure}
	\includegraphics[width=\linewidth]{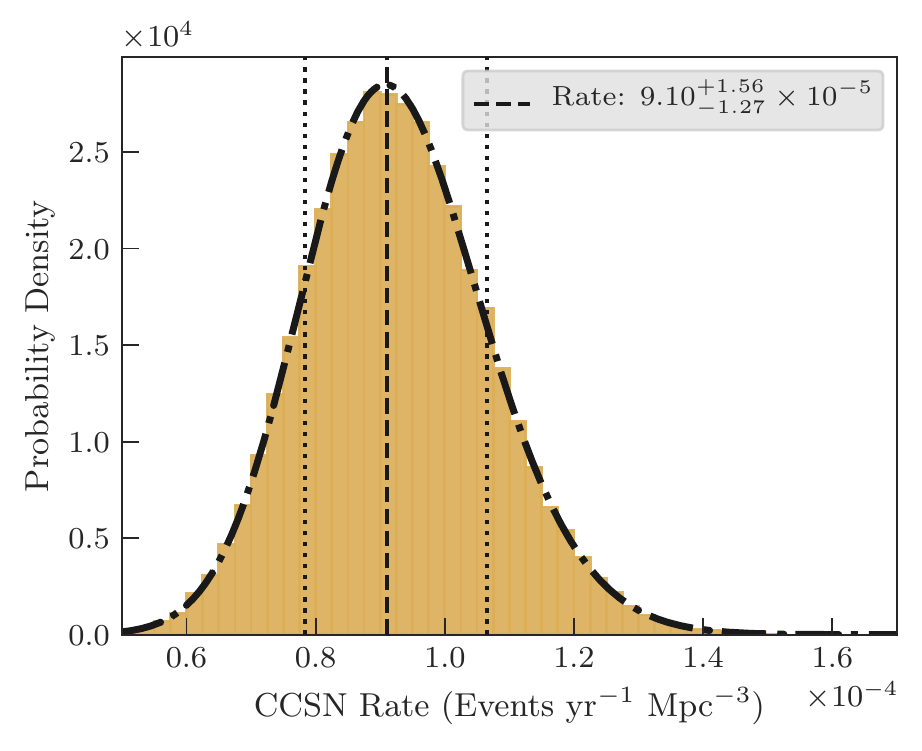}
    \caption{The probability distribution from our Monte-Carlo simulation of the intrinsic CCSN rate. This distribution includes both the statistical uncertainty associated with the observed number of CCSNe, and the survey systematics modelled in our simulation. The dot-dashed line is a skewed Gaussian fit to the data which is used to infer the maximum probability (dashed line) and the $1\sigma$ uncertainties (dotted lines).}
    \label{fig:ccRateDist}
\end{figure}

We now infer our final rate by bootstrap sampling from Equation~\ref{eqn:poiObs} and couple this with the output of the rate simulations. Firstly, we sample a value of $n$ from Equation~\ref{eqn:poiObs} to set the observed number of events for our trial. We then compare all outputs from the rate realizations described in Section~\ref{sec:ratesPTF} where $N_\mathrm{output} =  n$ to form a set of corresponding intrinsic rates $r_\mathrm{intrinsic}$. We then randomly select a value from this set to assume the intrinsic CCSNe rate for this realization. This process was repeated $\sim$10,000 times to build up a distribution of $r_\mathrm{intrinsic}$, which we present in Fig.~\ref{fig:ccRateDist}. A skewed Gaussian was then fit to the raw data to find the maximum probability (the mode) and the $1\sigma$ uncertainties around the median. We infer the volumetric rate of CCSNe to be $r^\mathrm{CC}_v=9.10_{-1.27}^{+1.56}\times10^{-5}\,\text{SNe yr}^{-1}\,\text{Mpc}^{-3}\, h_{70}^{3}$ at the simulated sample volume-weighted mean redshift of $ \langle z \rangle = 0.028$. The (16, 50, 84) percentiles from the distribution are $(7.83, 9.18, 10.66)\times10^{-5} \text{SNe yr}^{-1} \text{Mpc}^{-3}\, h_{70}^{3}$.

\subsubsection{Stripped-envelope supernova rate}
\label{SESN_Rate}

\begin{figure}
	\includegraphics[width=\linewidth]{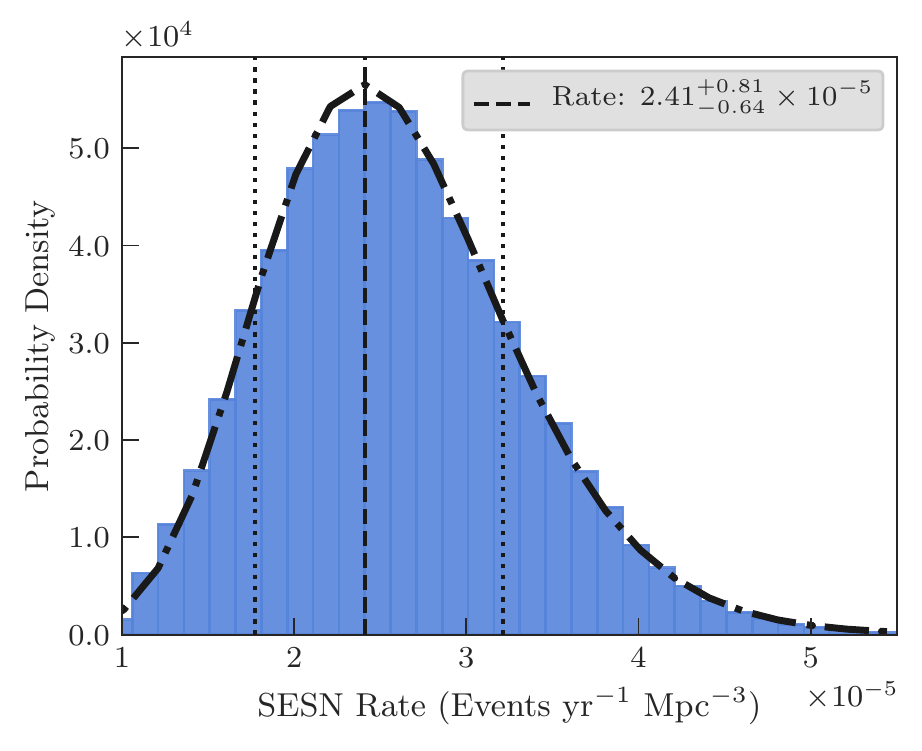}
    \caption{The intrinsic rate probability distribution for SESN sample. The distribution is formed by comparing the results of the Monte-Carlo simulation to the observed sample of SNe. The dot-dashed line represents the skewed Gaussian fit to the distribution from which the most probable intrinsic rate is inferred and shown by the dashed line. Our $1\sigma$ uncertainties around the mean are shown by the dashed lines.}
    \label{fig:SESNrateProbDist}
\end{figure}

We then applied our rate methodology to the SESNe subset of CCSNe from PTF. Our sample consists of 26 objects listed in Table \ref{tab:ccsn_properties}, including our two supernovae of unknown type. We thus adopt $N_\mathrm{obs}=26$ in Equation~\ref{eqn:poiObs} and follow the prescription of Section~\ref{sec:CCSN_Rate}. This measurement of the SESN rate is performed completely independently of the CCSN simulation and thus is unaffected by any assumptions on the SESN to CCSN population fraction. We perform 400 000 realizations of the SESN rate
, finding the most likely SESN rate to be $r^\mathrm{SE}_v=2.41_{-0.64}^{+0.81}\times10^{-5}\,\text{SNe yr}^{-1}\,\text{Mpc}^{-3}\, h_{70}^{3}$ at a volume weighted mean redshift of $ \langle z \rangle = 0.028$. Our functional fit to the data finds (16, 50, 84) percentiles of $(1.77, 2.43, 3.22)\times10^{-5}\, \text{SNe yr}^{-1}\,\text{Mpc}^{-3}\, h_{70}^{3}$. The probability distribution for this rate is presented in Fig.~\ref{fig:SESNrateProbDist}.

\subsection{Comparison to literature CCSN rates}
\label{sec:ccsn_lit}
We compare our CCSN rates to those from other surveys at $z<0.4$. We select works that have computed volumetric rates from rolling search data \citep[e.g.][]{Bazin2009}, or from those that have converted rates in units of host-galaxy properties to volumetric rates. We summarise the comparable rate measurements through increasing redshift in Table \ref{tab:ccsn_rates}.

\begin{figure*}
	\includegraphics[width=\linewidth]{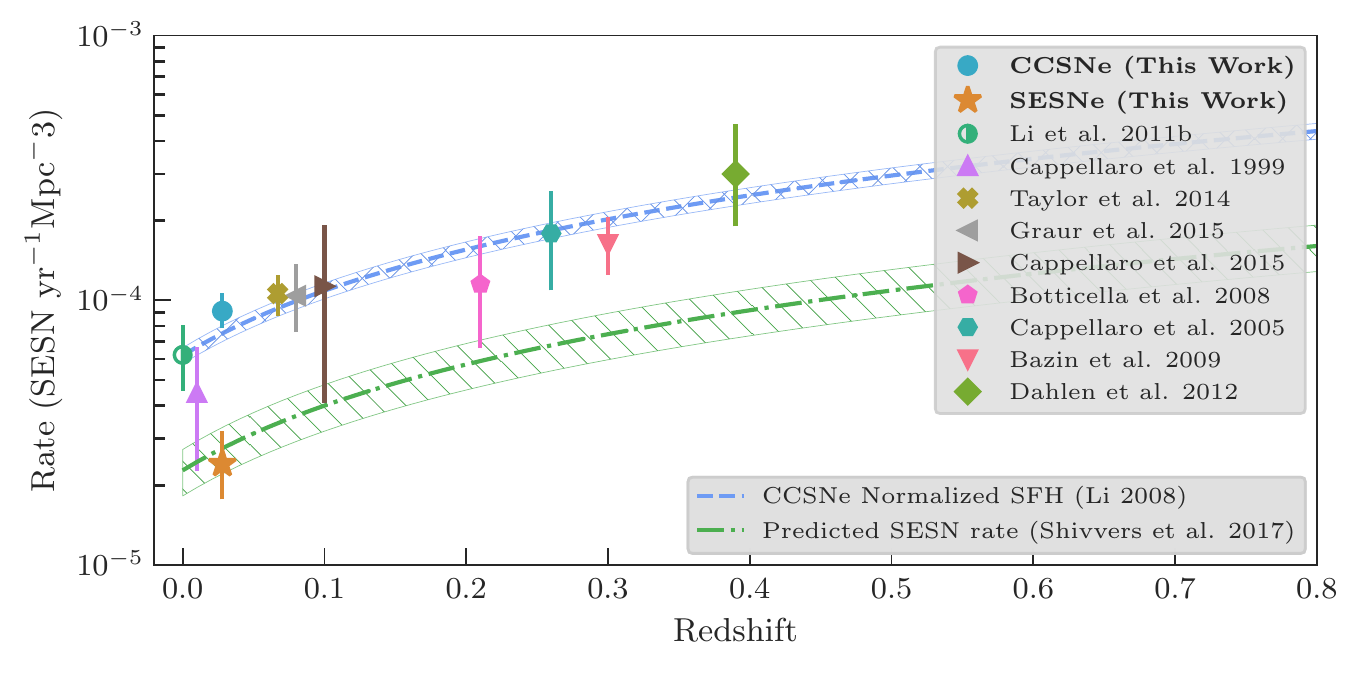}
    \caption{The volumetric CCSN rate evolution as a function of redshift compared to measurements from the literature \citep{Li2011_RateSize,Cappellaro1999,2014ApJ...792..135T,2015MNRAS.450..905G,Cappellaro15,Botticella2008,Cappellaro2005,Bazin2009,2012ApJ...757...70D}. We fit a normalised star-formation history \citep[blue dashed line;][]{2008MNRAS.388.1487L} to the CCSN rate sample, and represent our $1\sigma$ uncertainties with the hatched region. Our measurement of the SESN rate is the orange star, the SFH history we fit to the CCSN sample is scaled according to the relative population fractions of SESNe/CCSNe from \citet{2017PASP..129e4201S}, and is represented by the green dot-dashed line.}
    \label{fig:rates_CC_lit}
\end{figure*}

\begin{table*}
\begin{center}

    \caption{Literature CCSN rates, ordered by redshift. }
    \label{tab:ccsn_rates}
    \begin{tabular}{lllllll}
    \hline
    Survey & N$_{\mathrm{SN}}$ & Rate & Stat. uncertainty & Sys. Uncertainty & $z$ & Ref. \\
     & & $(\times10^{-4}\, \text{SNe yr}^{-1}\,\text{Mpc}^{-3})$ & &\\
    \hline
    
    LOSS  & 440 & $0.62$ & $\pm0.07$ & $_{-0.15}^{+0.17}$ & 0.00 & \citealt{Li2011_RateSize} \\
    
    -  & 67.4 & $0.43$ & $\pm0.17\, ^\mathrm{a}$ & - & 0.01 & \citealt{Cappellaro1999} \\
    
        PTF & 86 & 0.91 & $_{-0.13}^{+0.16}\, ^\mathrm{a}$ & - & 0.028 & This work \\
   
    SDSS & 89 & $1.06 $ & $\pm 0.19\, ^\mathrm{a}$& - & 0.072 & \citealt{2014ApJ...792..135T} \\
    
    SDSS & 16 & $1.04$ & $_{-0.26}^{+0.33}$ & $_{-0.11}^{+0.04}$  & 0.075 & \citealt{2015MNRAS.450..905G} \\
   
    SUDARE  & 50 & $1.13$ & $_{-0.53}^{+0.62}$ & $\pm 0.49$ & 0.10 & \citealt{Cappellaro15} \\
    
    STRESS & 44.95 & $1.15$ & $_{-0.33}^{+0.43} $& $_{-0.36}^{+0.42}$ & 0.21 & \citealt{Botticella2008} \\
    
    - & 32 & $2.2$ & $_{-0.7}^{+0.8}\, ^\mathrm{a}$ & - & 0.26 & \citealt{Cappellaro2005} \\
    
    SNLS & 117 & $1.42 $ &  $\pm 0.3$& $\pm 0.3$  & $0.30$ & \citealt{Bazin2009} \\
    
    HST & 9& $3.00$ & $_{-0.94}^{+1.28}$ & $_{-0.57}^{+1.04}$ & $0.39$ & \citealt{2012ApJ...757...70D} \\
    \hline
    
    $^\mathrm{a}$ Total quoted uncertainty.
		
    \end{tabular}

        \end{center}
    
    \end{table*}

We present this combined literature sample, including our CCSN and SESN rate measurements, in Fig.~\ref{fig:rates_CC_lit}, correcting to our assumed cosmology where relevant. The increasing CCSN rate with redshift reflects the increasing SFR with look back time - which CCSNe are excellent probes of, given that they explode relatively promptly after formation ($\lesssim40\,\mathrm{Myr}$). We normalise the cosmic Star Formation History (SFH) as parameterised by \citet{2008MNRAS.388.1487L} with the \citet{2001MNRAS.326..255C} functional form (blue dashed line in Fig.~\ref{fig:rates_CC_lit}). The CCSN rate follows this simple SFH which, from $z=0$ over a look back time of $\sim4.3$ Gyr, increases by a factor of $\sim 4.2$.

We are unable to make any assertions on the fraction of CCSN that are SESNe from our data as this ratio is an implicit assumption used in the CCSN rate simulation. However, we can use the \citet{2017PASP..129e4201S} population fractions to probe any systematic biases in our simulation pipeline. It is natural to expect that the brighter intrinsic luminosities of SESNe would make them easier to discover and, hence, introduce a malmquist bias. We explore this by taking our normalised SFH and scaling it by the estimated SESN/CCSN population fractions from \cite{2017PASP..129e4201S}. This is presented in Fig.~\ref{fig:rates_CC_lit} by the green dot-dashed line and is in good agreement with our measured SESN rate represented by the orange star. We are, therefore, confident that our rate simulations are unaffected by the malmquist bias and further adds credence to our independent measurement of the absolute SESN rate.

\subsection{SLSN-I Rate}
\label{sec:SLSN_Rate}

We now examine the rate of SLSNe-I from the PTF data. In principle, the method we follow here is identical to the CCSN rate measurements but with all PTF fields included in the simulation. This change is justified as; i) the longer duration of a typical SLSN light curve makes the detection of such an event less sensitive to the cadence of any particular PTF field, ii) our SLSN-I sample is small to begin with and prioritizing cadence over sky area significantly affects this statistics limited sample. These additional fields and SN locations are marked in Fig.~\ref{fig:ptfSky}. 
\begin{figure}
	\includegraphics[width=\linewidth]{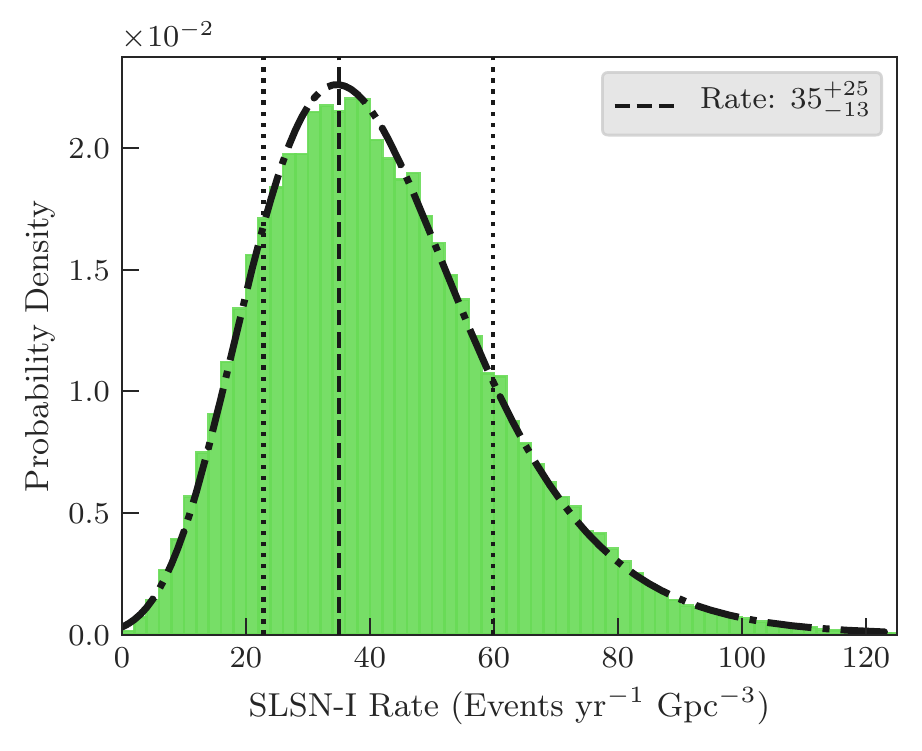}
    \caption{The SLSN-I rate probability distribution as inferred from our Monte-Carlo simulation. We include both statistical and systematic uncertainties within our simulation results and show a log-normal functional fit (dot-dashed line) to the distribution which was used to find the most probable intrinsic rate (dashed line). The $1\sigma$ uncertainties are calculated from the area containing 68.3 per cent of the probability around the median which is bound by the dotted-line.}
    \label{fig:SLSNrateProbDist}
\end{figure}

The rate simulations were conducted using the methodology presented in Section~\ref{sec:ratesPTF}, with Poisson variation in the observed sample accounted for following Equation~\ref{eqn:poiObs}, and template light curves generated with the prescription of Section~\ref{sec:simSLSN}. Again, in comparison to the CCSN rates, the variables $T_\mathrm{obs}$ and $V_\mathrm{obs}$ are adjusted to encompass the entire PTF visible sky from 2010--2012 out to a redshift of $z=0.2$. 
We performed $1\times10^6$ realizations of the SLSNe-I rate and compared the simulation output observations to the Poisson distributed observed sample of SLSNe-I in PTF with $N_\mathrm{obs}=8$. The probability distribution of rates, calculated following the bootstrap method in Section~\ref{sec:CCSN_Rate}, is shown in Fig.~\ref{fig:SLSNrateProbDist} and from the functional fit to the data we infer the SLSN-I rate to be $r^\mathrm{SLSN-I}_v=35_{-13}^{+25}\, \text{SNe yr}^{-1}\text{Gpc}^{-3}\, h_{70}^{3}$ and (16, 50, 84) percentiles of $(22, 39, 60)\, \text{SNe yr}^{-1}\text{Gpc}^{-3}\, h_{70}^{3}$.

\subsection{Comparison to literature SLSN rates}

There are few well-measured SLSN-I rates in the literature as a result of the low intrinsic volumetric rate and relatively recent shift to wide-area un-targeted rolling sky surveys. This is further complicated by uncertainties in the classification of SLSNe-I; either due to insufficient wavelength coverage to determine the H-poor/H-rich subtypes, or poor signal-to-noise making it difficult to distinguish the faint absorption features which mark these events at optical wavelengths. This is only exacerbated at higher redshifts where high signal-to-noise spectra are expensive to collect.

\begin{figure*}
	\includegraphics[width=\linewidth]{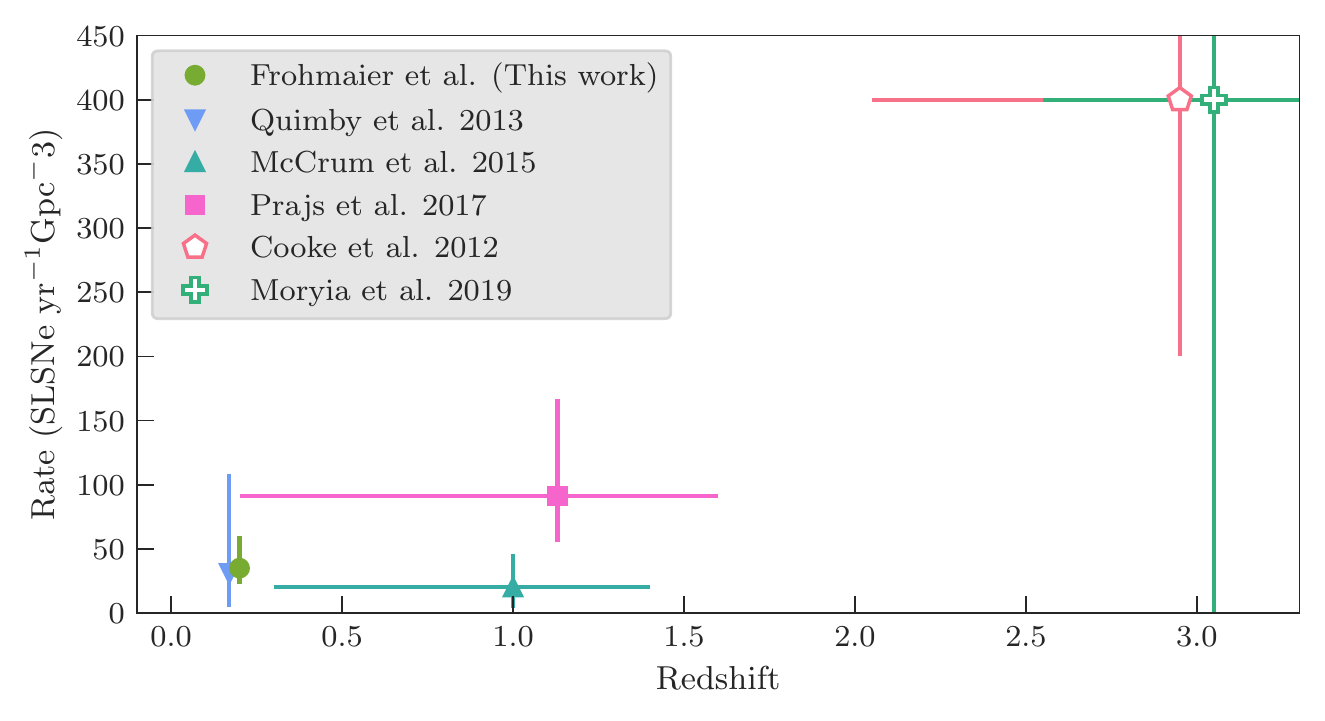}
    \caption{A literature comparison of the volumetric rates of SLSNe. The open symbols denote rate measurements that do not distinguish between hydrogen-poor SLSNe and objects that show evidence for hydrogen in their spectra. We have offset some rate measurements slightly in redshift for clarity. }
    \label{fig:rates_SLSN_Lit}
\end{figure*}

\begin{table*}
\begin{center}

    \caption{Literature SLSN-I rates, ordered by redshift. The SLSN sample for the high-$z$ rates of \citet{Cooke2012} and \citet{2019ApJS..241...16M} are not split into sub-types.
    }
    \label{tab:slsn_rates}
    
    \begin{tabular}{llllll}
        \hline
    Survey & N$_{\mathrm{SN}}$ & Rate & Uncertainty & $z$ & Ref. \\
    & &  $(\text{SNe yr}^{-1} \text{Gpc}^{-3})$ & & & \\
        \hline

ROTSE-IIb & 1 & $32$&$_{-26}^{+77}$ & $0.17$& \citealt{Quimby2013} \\
PTF &8 &$35$&$_{-13}^{+25}$ &$0.17$& This work. \\
PS1 &$<10^{+4}_{-3}$ & $3 - 8 \times 10^{-5}$& - & $0.3\le z\le 1.4$& \citealt{McCrum2015} \\
 & & of CCSN rate & & &\\
SNLS &3& $91$&$_{-36}^{+76}$ & 1.13 &\citealt{Prajs2016} \\
CHFT Legacy Survey Deep Fields &$2\, ^{\mathrm{a}}$&$\sim400$& - & $2.0\le z\le 4.0$ & \citealt{Cooke2012} \\
 Subaru HIgh-Z sUpernova CAmpaign &$>1\, ^\mathrm{a}$&$\sim 400$& $\pm 400$ & $2.5 \le z \le 3.5$ &\citealt{2019ApJS..241...16M} \\
    \hline
    $^\mathrm{a}$ Rate not split into SLSN sub-types\\
    \end{tabular}

        \end{center}
    \end{table*}

In Fig.~\ref{fig:rates_SLSN_Lit} we show the literature SLSN rate measurements and briefly present these results in increasing redshift order in Table \ref{tab:slsn_rates}. As with CCSNe, we see an increasing trend in SN rate with redshift. However, given that there are only a handful of poorly constrained high-$z$ rates in the literature, it is difficult to determine how closely the rate follows star formation rate evolution with cosmic time, or whether other astrophysical factors become more dominant in the early Universe.

\section{Host stellar mass rates}
\label{sec:host_mass_rates}
Our analysis can be extended to look at the relative and absolute rates of supernova populations as a function of the host-galaxy stellar mass. This will allow us to examine whether the volumetric rate ratios hold, or if different SN sub-types preferentially occur in galaxies of different masses.

The volumetric rates framework, described in Section~\ref{sec:ratesPTF}, can be modified to produce rates as a function of stellar mass ($\text{SNe yr}^{-1}\,\text{M}_\odot^{-1}$)  through the use of a galaxy stellar mass function \citep[GSMF; e.g.][]{2001MNRAS.326..255C,2003ApJ...585L.117B}. The GSMF is the number density (in units of $\mathrm{Mpc}^{-3}$) of galaxies in a given logarithmic stellar-mass interval and is typically described by a double Schechter function \citep{Schechter1976}. In Section~\ref{sec:CCSample}, we established 9 fixed-volume sub-surveys for our rate calculation, we use these volumes to convert the GSMF from a number density to the raw number of galaxies and then integrate over any given mass range to calculate the total stellar mass in that volume. We adopt the double Schechter function fit of the \citet{2017MNRAS.470..283W} GSMF for our analysis.

It is now possible to continue with our rate calculation by drawing random values for the intrinsic rate per unit mass and multiply this by the total stellar mass from the GSMF to produce a simulated number of SNe. This is analogous to Equation~\ref{eqn:poiSim}, but with a change of variable from $V_\mathrm{obs} \rightarrow \mathrm{M}_\mathrm{*}$, where $\mathrm{M}_\mathrm{*}$ is the total stellar mass in our bin. This simple change to our pipeline allows us to perform realizations of the rate per unit mass to compare the simulated population to the observed population of SNe. Once again, the simulated intrinsic rates that reproduce the observed population determine our measurement of the rate.

The final piece of our puzzle requires us to measure the host galaxy masses for all the CCSNe and SLSNe in our sample. In \citet[][]{Perley2016} host galaxy masses were presented for most of our SLSN sample, however, owing to the different luminosity cuts between their and the \citet{DeCia2018} SLSN sample, we are missing the galaxy masses for PTF12gty and PTF12hni; a description of how we measure galaxy masses for all objects follows. A consequence of our low-redshift CCSN sample is that many of our host-galaxies are in SDSS fields. For galaxies with a spectroscopic redshift, an SED was constructed from stellar population modelling \citep{2005MNRAS.362..799M} and fit to the $ugriz$ photometry fixed at that redshift. We used the SDSS computed \citep{2006ApJ...652...85M} stellar masses where available. The galaxy masses for all our CCSNe and additional SLSNe-I are listed in Table~\ref{tab:host_masses}.

In the event a SN host in our sample was not in an SDSS field or did not have spectro-photometric derived properties, we perform our own SED template fits fixed at the SN redshift. Our galaxy photometry was obtained primarily from SDSS, but if the target field was not covered we turned to the PanSTARRS catalogues \citep{2016arXiv161205243F}. We used the galaxy fluxes from apertures determined by the \texttt{cModelMag} or \texttt{KronFlux} for the $griz$ photometry from SDSS and PanSTARRS catalogues respectively. We next adopt the P\'EGASE.2 spectral synthesis code \citep{1997A&A...326..950F,2002A&A...386..446L} to construct the SEDs for our galaxies. The templates SEDs are fit to the observed fluxes through a $\chi^{2}$ minimisation; this method is extensively described in \citet{2010MNRAS.406..782S,2018ApJ...854...24K,2020MNRAS.494.4426S}. From the template fits, we derive our sample host galaxy masses and list them in Table~\ref{tab:host_masses}.

\subsection{CCSN and SESN galaxy mass rates}

To calculate the CCSN rate as function of galaxy stellar mass, we first split our sample up into 3 mass bins: $\le 9.5\, \textrm{log}\, M_{*}$ (low), $9.5 < \textrm{log}\, M_{*} \le 10.5$ (intermediate), and  $10.5 < \textrm{log}\, M_{*} \le 11$ (high). The number of SNe in each bin determines the $N_\mathrm{obs}$ parameter described in Equation~\ref{eqn:poiObs}. In each bin we perform 100 000 realizations of the mass-rate using the methods described in this and previous Sections. We then compare the output of the simulation to the observed number of SNe by PTF to connect back to the intrinsic rate distribution. This analysis was conducted following the methods presented in Section~\ref{sec:rates}. For each mass bin a rate probability distribution was inferred, with the 16, 50, and 84 percentiles representing our median rate and $1\sigma$ uncertainties. These results are presented in Table~\ref{tab:all_host_rates} in units of SNuM\footnote{The SNuM is the supernova rate per unit mass, defined as the rate of SNe per century per $10^{10}\mathrm{M}_\odot$ or equivalently, $1\times 10^{-12}\mathrm{yr}^{-1}\mathrm{M}_\odot^{-1}$.}.

\begin{table}
\begin{center}
    \caption{The rates of CCSNe, SESNe, and SLSNe-I as a function of the host galaxy stellar mass. The SESNe results are shown for a sample including SN IIb and, in parentheses, excluding SN IIb. This was done to allow a comparison with the host galaxy rates of \citet{2017ApJ...837..120G}, who excluded SN IIb from their sample.}
    \label{tab:all_host_rates}
\begin{tabular}{lccc} 
\hline
SN Type              & Mass    & N$_\mathrm{SN}$ & Rate                  \\
                     & log $M_{*}(\mathrm{M}_\odot$) &     & (SNuM)   \\ 
\hline
                     & $\le9.5$     &  48   & $2.83_{-0.54}^{+0.60}$                 \\
CCSNe                & $9.5 < \textrm{log} M_* \le 10.5$     &   26  & $0.44_{-0.11}^{+0.13}$                  \\
                     & $10.5 < \textrm{log} M_* \le 11$      &  12   & $0.16_{-0.06}^{+0.07}$                  \\ 
\hline
                     & $\le9.5$      &  10 (8)   & $0.58_{-0.22}^{+0.28}$ $\left(0.48_{-0.20}^{+0.25}\right)$           \\
SESNe  & $9.5 < \textrm{log} M_* \le 10.5$     &  10 (8)   & $0.16_{-0.06}^{+0.08}$ $\left(0.13_{-0.05}^{+0.07}\right)$           \\
                     & $10.5 < \textrm{log} M_* \le 11$      &   6 (5)  & $0.08_{-0.04}^{+0.05}$ $\left(0.07_{-0.03}^{+0.05}\right)$         \\ 
\hline
SLSNe                & $\le9.5$       &  8   & $1.73_{-0.65}^{+1.17} \times 10^{-3}$            \\
    \hline
                   
\end{tabular}\\
\end{center}
\end{table}

\begin{figure}
	\includegraphics[width=\linewidth]{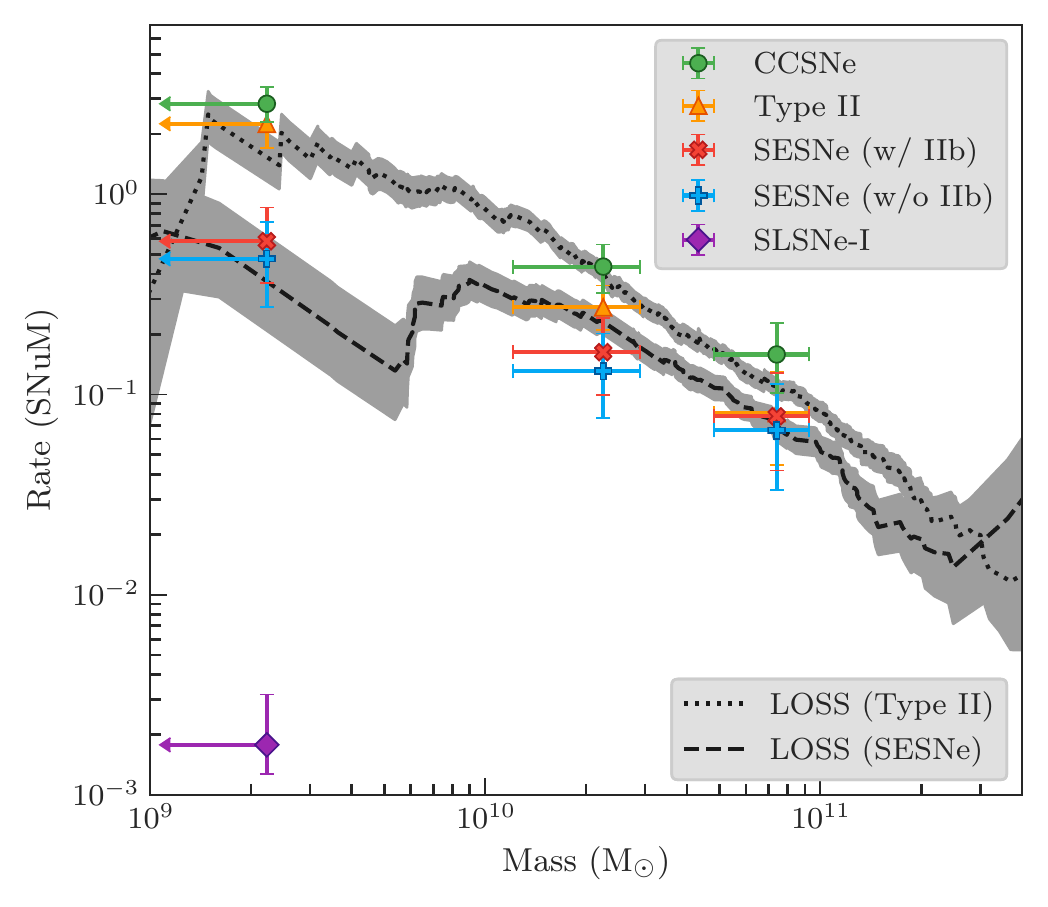}
    \caption{The SN rates as a function of the host galaxy stellar mass for all our SN sub-types. The error bars show the $1\sigma$ uncertainties on the measured rates on the vertical, and the interval containing 68.3 per cent of the simulated galaxy masses around the mass-weighted mean on the horizontal. The CCSN rate is denoted by the green circles. The SESNe was calculated including SN IIb in the sample (red cross) and excluding SN IIb (blue plus) in order to make a direct comparison to the LOSS reanalysis \citep{2017ApJ...837..120G} which did not include SN IIb in its SESN sample. The SLSN-I rate is presented by the purple diamond. We also show the rates in the LOSS reanalysis for their SN II sample (dotted line) and their SESN sample (dashed line), the grey shaded region denote the $1\sigma$ uncertainty on their measurement. We do not directly measure the SN II rate, but instead infer it as the difference between the CCSN and SESN rates, allowing for a comparison to LOSS.}
    \label{fig:rateHostMass}
\end{figure}

\subsection{SLSN-I galaxy mass rate}

Several studies have shown that SLSNe-I are predominantly found in low-mass galaxies \citep[e.g.][]{2014ApJ...787..138L,Perley2016,2017MNRAS.470.3566C}. For our analysis of SLSN rates, we extend the PTF sample in \citet{Perley2016} to include PTF12gty and PTF12hni from the \citet{DeCia2018} analysis. The host of PTF12gty was not detected in optical imaging from SDSS. To estimate an upper-limit on the host stellar mass, we fixed the template SEDs at the redshift of the SN and assumed an SDSS limiting magnitude of m$_r = 23.1$. We place an upper-limit on the host stellar mass at $\mathrm{log\, M}_{*} < 9.5 \mathrm{M}_\odot$. Considering all other galaxies in the SLSN-I sample are below this limit, we perform a simulation of the SLSN-I rate with host galaxies drawn from the GSMF with $\mathrm{log\, M}_{*} < 9.5 \mathrm{M}_\odot$. We performed 100 000 realizations of the rate and matched the output SLSN-I populations to the PTF observed SLSN-I population following the methods described in previous sections. For the simulations that reproduce the observed sample, we infer the galaxy stellar mass rate as the median of the rate probability distribution and quote the 16 and 84 percentiles as our $1\sigma$ uncertainties. These results are shown in Table~\ref{tab:all_host_rates}.

\section{The relative rates of CCSNe and SLSNe-I}
\label{sec:relative}

We now use our rate estimates to examine the relative rates of CCSNe to SLSNe-I in the PTF sample. We choose to only use the results of this work to mitigate biases introduced by different survey selection effects and systematic errors.

The relative rates of SNe are a useful diagnostic in constraining progenitor scenarios for different SN types and/or the role of galaxy properties in hosting these SNe. Under the assumption of a constant Initial Mass Function (IMF), the evolution of relative rates of SESN-to-CCSNe-to-SLSNe may help break the degeneracy between the host star formation rate and metallicity for core collapse progenitors (for instance, if the ratios remains constant with redshift, this would heavily imply that the production of the respective progenitors were mostly reliant on the local star-forming properties, whilst an evolving ratio would suggest that one progenitor may be more dependant upon other environmental properties such as metallicity). However, this requires good estimates of how relative rates evolve across all redshifts \textit{and} reliable progenitor models for all SN subtypes. In this regard, we are currently limited in what can be achieved with rates measured from a single survey at a single redshift, but we begin the efforts by analysing our low-redshift sample. 

\subsection{Relative Volumetric Rate}

\begin{figure}
	\includegraphics[width=\linewidth]{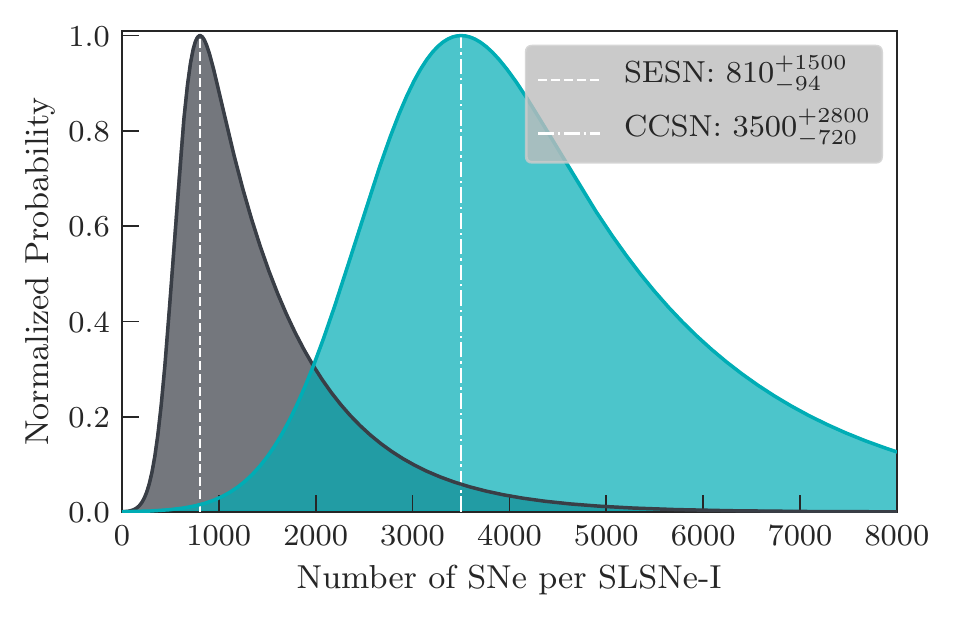}
    \caption{The probability distributions for the expected number of SLSNe-I per CCSN (dot-dashed line) and SESN (dashed line) from the results of our volumetric rate simulation.}
    \label{fig:compare2SLSN}
\end{figure}

To measure the relative rates of CCSNe-to-SLSNe, we first scale our CCSN rate to the redshift of our SLSN measurement using the normalized SFH in Section~\ref{sec:ccsn_lit}. We then perform a Monte Carlo simulation by drawing random values for the rates of CCSNe and SLSNe from the distributions show in Figs.~\ref{fig:ccRateDist} and~\ref{fig:SLSNrateProbDist} respectively. In total, $1\times10^5$ realizations were made to find the number of CCSNe per SLSN-I. The resulting probability distribution is shown in Fig.~\ref{fig:compare2SLSN} and from this we infer the SLSN-I--to--CCSN fraction to be $1/3500^{+2800}_{-720}$. The process was repeated in an identical fashion for the SESN sub-sample, from which we infer a SLSN-I--to--SESN fraction to be  $1/810^{+1500}_{-94}$ and, again, present the probability distribution in Fig.~\ref{fig:compare2SLSN}. \citet{Quimby2013} estimated that for every one SLSN-I there are 1000 -- 20 000 CCSNe. This is consistent with our results; however, our larger SLSN-I sample size offers improved constraints on the population fraction.

\subsection{Relative Host Galaxy Rates}

We now compare the relative fractions of SLSNe-I to only the CCSNe and SESNe that were hosted in galaxies of similar stellar masses. For each SN sub-type, we randomly sample the full rate probability distribution for each of the host galaxy rates. Since we only have SLSNe-I in our lowest mass bins we only present results for the relative rates in galaxies with $\mathrm{log\, M}_{*} < 9.5 \mathrm{M}_\odot$. For each trial rate from the SN class, we compare their relative frequencies and repeat this process 100 000 times. Our analysis infers that the SLSN-I--to--CCSN fraction is $1/1700^{+1800}_{-720}$, and a SLSN-I--to--SESN fraction to be $1/300^{+380}_{-170}$.

We next check whether the dearth of SLSN-I in higher-mass galaxies is likely an astrophysical effect, or a consequence of their relative rarity conflated with the PTF observing strategy. To test this, we assume the relative frequency between SESNe and SLSNe-I in the low-mass bin remains constant across all host galaxies and scale the SESN rate accordingly. Using this scaled rate as our fiducial SLSN-I rate, we perform a simulation of the PTF survey, drawing SN host galaxies from the GSMF in the range $9.5 < \textrm{log M}_{*} \le 10.5$. The simulated light curves are computed through the analysis framework described in Section~\ref{sec:rates} to assess how many, if any, SLSNe-I would have been discovered in higher-mass hosts. From our analysis, if the SESN-to-SLSN-I ratio holds then the Poisson probability of observing zero events is ${<}7.5\times10^{-6}$. We extend the calculation to include all galaxies with $\textrm{log M}_{*} > 9.5\textrm{M}_\odot$ and find the Poisson probability of PTF discovering zero SLSNe-I to be ${<}1.1\times10^{-9}$. Given this result, we confirm previous reports \citep[e.g.][]{Leloudas2015,Angus2016,2018MNRAS.473.1258S} that SLSNe-I show a preference for low-mass (low-metallicity) environments.\\

These relative fractions could be used to infer the potential ranges of progenitor masses for the different SN classes. However, without a firm knowledge of the underlying IMF representing each transient population, this becomes difficult. The functional form of the IMF may be altered by both the stellar metallicity and the binarity of the population.  Additionally, factors such as the level of stellar rotation and magnetic field strength can also alter the evolution of massive stars. This changes the amount of stellar material lost to winds between main-sequence and collapse,
thus weakening the assumption that the natal IMF is representative of the mass function at the point of explosion.

\section{Summary}

We have presented the volumetric rates for three types of SNe occurring from massive star progenitors -- core collapse supernovae (CCSNe), stripped-envelope supernovae (SESNe), and hydrogen-poor superluminous supernovae (SLSNe-I). Our data are from the PTF and our rate calculation methodology builds on the detection efficiencies of \citetalias{Frohmaier17}. To measure the volumetric rates for CCSN populations, we adopted the light curve templates of \citet{Vincenzi2019}, and coupled them to a statistically representative simulation of PTF. Our sample of CCSNe consists of 84 spectroscopically classified objects and 2 photometrically identified candidates across the II, IIb, IIn, and Ibc sub-classes. Our simulations infer that, at a mean redshift of $ \langle z \rangle = 0.028$, the CCSN rate is
\begin{equation*}
  r^\mathrm{CC}_v=9.10_{-1.27}^{+1.56}\times10^{-5}\,\text{SNe yr}^{-1}\,\text{Mpc}^{-3}\, h_{70}^{3}.
\end{equation*}

We then performed an independent rates analysis on the sub-sample of CCSNe that do not show strong hydrogen in their spectra - SESNe. The principles behind the methodology were identical to the CCSNe, except we restricted ourselves to the SN Ib/c and IIb templates and worked with a sample of 26 objects: 24 spectroscopically classified and 2 photometrically identified candidates. From this we measure the SESN rate to be
\begin{equation*}
  r^\mathrm{SE}_v=2.41_{-0.64}^{+0.81}\times10^{-5}\, \text{SNe yr}^{-1}\,\text{Mpc}^{-3}\, h_{70}^{3}.
\end{equation*}

Including our measurements in a collection of other CCSN results from the literature to a redshift of $z=0.4$ demonstrates that the CCSN rate traces the cosmic SFH. This translates to an increase of $\sim4.2\times$ in both the CCSN rate and star formation rate over the past $4.3$ Gyr of cosmic evolution.

The population fraction of CCSNe obscured by large amounts of dust may be as high as $\sim20\%$ in local Universe optical surveys, \citep{Mattila2012}, particularly hosts with intrinsically high levels of dust, such as Luminous Infrared Galaxies. While our core collapse light curve models are extincted following the \citet{Vincenzi2019} population, our simulations do not account for a completely obscured population. If this effect is truly significant, then the intrinsic CCSN rate may be higher than reported in this work. 

We investigated the rate of SLSNe-I from the PTF sample \citep{Quimby2018,DeCia2018}. We performed our Monte-Carlo simulation on the entire PTF sky and adopted a light curve model for a transient powered by the spin-down of a magnetar. We performed $1\times10^6$ rate realizations and compared the output to our observed sample at a mean redshift of $\langle z \rangle =0.17$. From this, we infer the SLSN-I rate to be
\begin{equation*}
  r^\mathrm{SLSN-I}_v=35_{-13}^{+25}\, \text{SNe yr}^{-1}\text{Gpc}^{-3}\, h_{70}^{3}.
\end{equation*}

Finally, we compared the relatives fractions of CCSNe to SLSNe from the perspective of both the co-moving volume of the universe and their host galaxy environments. For the volumetric rates, we find one SLSN-I per $\sim 3500^{+2800}_{-720}$ CCSN and one SLSN-I per $810^{+1500}_{-94}$ SESN. When we restrict our galaxy sample to hosts with $\mathrm{log\, M}_{*} < 9.5 \mathrm{M}_\odot$, we find one SLSN-I per $1700^{+1800}_{-720}$ CCSN and one SLSN-I per $300^{+380}_{-170}$ SESN. The propensity for SLSNe to occur in low-mass galaxies, when compared to the volumetric rate ratios, suggests a progenitor scenario favouring these environments. 

In the coming decade, the next generation of wide-fast-deep surveys, such as the Rubin Observatory Legacy Survey of Space and Time \citep[LSST;][]{LSSTREF}, will discover transients in vastly greater quantities than we present today. This is will allow measurements of SN rates across multiple redshift bins out to large distances from a single survey, and thus, will not suffer differing survey systematics as current literature comparisons do. Measuring rates for multiple SN classes is key to understanding how both SN progenitors and their host systems evolve with cosmic time. The methods presented in this paper will be applicable to rate calculations from LSST and other sky surveys.

\section*{Acknowledgements}

We thank the anonymous referee for their useful comments. We are grateful to Laura Nuttall and Or Graur for their support and scientific discussions during the preparation of this manuscript.

We acknowledge support from EU/FP7 ERC grant number 615929 and the Science \& Technology Facilities Council (STFC) grant number ST/P006760/1 through the DISCnet Centre for Doctoral Training. We acknowledge the use of the IRIDIS High Performance Computing Facility, and associated support services, at the University of Southampton. This research used resources of the National Energy Research Scientific Computing Center, a DOE Office of Science User Facility supported by the Office of Science of the U.S. Department of Energy under Contract No. DE-AC02-05CH11231. Observations were obtained with the Samuel Oschin Telescope and the 60-inch Telescope at the Palomar Observatory as part of the Palomar Transient Factory project, a scientific collaboration between the California Institute of Technology, Columbia University, Las Cumbres Observatory, the Lawrence Berkeley National Laboratory, the National Energy Research Scientific Computing Center, the University of Oxford, and the Weizmann Institute of Science. Funding for the Sloan Digital Sky Survey IV has been provided by the Alfred P. Sloan Foundation, the U.S. Department of Energy Office of Science, and the Participating Institutions. SDSS-IV acknowledges
support and resources from the Center for High-Performance Computing at
the University of Utah. The SDSS web site is www.sdss.org.

SDSS-IV is managed by the Astrophysical Research Consortium for the 
Participating Institutions of the SDSS Collaboration including the 
Brazilian Participation Group, the Carnegie Institution for Science, 
Carnegie Mellon University, the Chilean Participation Group, the French Participation Group, Harvard-Smithsonian Center for Astrophysics, 
Instituto de Astrof\'isica de Canarias, The Johns Hopkins University, Kavli Institute for the Physics and Mathematics of the Universe (IPMU) / 
University of Tokyo, the Korean Participation Group, Lawrence Berkeley National Laboratory, 
Leibniz Institut f\"ur Astrophysik Potsdam (AIP),  
Max-Planck-Institut f\"ur Astronomie (MPIA Heidelberg), 
Max-Planck-Institut f\"ur Astrophysik (MPA Garching), 
Max-Planck-Institut f\"ur Extraterrestrische Physik (MPE), 
National Astronomical Observatories of China, New Mexico State University, 
New York University, University of Notre Dame, 
Observat\'ario Nacional / MCTI, The Ohio State University, 
Pennsylvania State University, Shanghai Astronomical Observatory, 
United Kingdom Participation Group,
Universidad Nacional Aut\'onoma de M\'exico, University of Arizona, 
University of Colorado Boulder, University of Oxford, University of Portsmouth, 
University of Utah, University of Virginia, University of Washington, University of Wisconsin, 
Vanderbilt University, and Yale University.

\section*{Data Availability}
The data underlying this article will be shared on reasonable request to the corresponding author.







\appendix

\section{Sub-survey Parameters}
Table~\ref{tab:lc_sky_area_nights} shows the parameters for each of the sub-surveys we construct for the simulations in Section~\ref{sec:CCSN_Rate}.
\begin{table}
\centering
\caption{The 9 sky areas we use for our CCSN rate measurements and survey simulations. The R.A. and decl. values are the central coordinates for each of the sub-survey's total footprint area. The start and end dates denote the time period over which a SN must have exploded in for both the real observations and simulations.}
\label{tab:lc_sky_area_nights}
\begin{tabular}{cccrrr}
\hline
Sub-survey & R.A. & decl. &  Area & Start & End\\
 &(J2000) &(J2000) &(deg$^2$)& (MJD) & (MJD)\\
\hline
2010 A & 12:46:00 & +35:00:00 & 10155 & 55257 & 55397  \\
2010 B & 22:20:00 & +06:30:00 & 2716 & 55347 & 55517 \\
2010 C & 02:30:00 & +06:30:00 & 4074 & 55412 & 55517 \\
2011 A & 15:10:00 & +40:00:00 & 3524 & 55627 & 55722 \\
2011 B & 17:10:00 & +50:00:00 & 1289 & 55717 & 55857 \\
2011 C & 22:00:00 & +10:00:00 & 3385 & 55742 & 55857 \\
2012 A & 14:20:00 & +45:00:00 & 4456 & 56007 & 56077 \\
2012 B & 16:00:00 & +20:00:00 & 1841 & 56077 & 56152 \\
2012 C & 23:55:00 & +00:00:00 & 314 & 56107 & 56192  \\
\hline
\end{tabular}
\end{table}

\section{Galaxy Stellar Masses}
Table~\ref{tab:host_masses} lists the CC and SLSNe-I host galaxy stellar masses used in Section~\ref{sec:host_mass_rates}.
\begin{table*}
\caption{The host galaxy masses used to measure our rate per unit mass. We show all CCSNe, along with two SLSNe-I that were excluded by \citet{Perley2016} luminosity cut. Our masses were obtained from SDSS data products \citep[methods described in][]{2006ApJ...652...85M}, or from our own SED template fitting routines to either SDSS or PanSTARRS photometry.}
    \label{tab:host_masses}
    \begin{center}
    \begin{tabular}{llllll}
        \hline
PTF Name & Redshift & SN Type & Galaxy Mass & Uncertainty & Source \\
 & & & $\mathrm{log}\, \mathrm{M}_{*} (\mathrm{M}_\odot)$ & & \\
\hline
10acbu & 0.0099 & SN Ic & 10.78 & $_{-0.09}^{+0.07}$ & SDSS \\
10bau & 0.026 & SN IIP & 10.03 & $_{-0.11}^{+0.36}$ & SDSS \\
10bgl & 0.03 & SN IIP & 10.58 & $\pm0.02$ & SDSS Phot \\
10bld & 0.027 & SN II & 8.95 & $_{-0.52}^{+0.06}$ & SDSS \\
10cxx & 0.034 & SN IIb & 9.58 & $_{-0.00}^{+0.37}$ & SDSS \\
10dvb & 0.023 & SN IIP & 10.04 & $_{-0.33}^{+0.00}$ & SDSS \\
10ehy & 0.006 & SN IIP & 7.81 & $_{-0.04}^{+0.14}$ & SDSS \\
10eqe & 0.031 & SN II & 8.75 & $_{-0.20}^{+0.13}$ & SDSS \\
10eqi & 0.03 & SN Ib & 10.86 & $_{-0.43}^{+0.00}$ & SDSS \\
10eqz & 0.029 & SN II & 9.47 & $\pm0.03$ & PanSTARRS Phot \\
10feq & 0.028 & SN Ib & 9.75 & $_{-0.09}^{+0.19}$ & SDSS \\
10fgz & 0.012 & SN II & 8.57 & $\pm0.02$ & SDSS Phot \\
10fjh & 0.032 & SN IIn & 10.88 & $\pm0.04$ & SDSS Phot \\
10fmr & 0.02 & SN IIb & 8.63 & $_{-0.00}^{+0.61}$ & SDSS \\
10fqg & 0.028 & SN IIb & 8.32 & $_{-0.16}^{+0.00}$ & SDSS \\
10fqx & 0.017 & SN IIP & 9.34 & $\pm0.01$ & SDSS Phot \\
10gki & 0.026 & SN II & 10.85 & $_{-0.11}^{+0.18}$ & SDSS \\
10gva & 0.028 & SN II & 8.36 & $_{-0.04}^{+0.14}$ & SDSS \\
10gxi & 0.029 & SN II & 8.52 & $_{-0.03}^{+0.02}$ & SDSS \\
10hlc & 0.035 & SN II & 8.34 & $_{-0.00}^{+0.22}$ & SDSS \\
10hny & 0.027 & SN IIP & 10.10 & $_{-0.09}^{+0.23}$ & SDSS \\
10hpa & 0.03 & SN IIP & 8.98 & $_{-0.23}^{+0.00}$ & SDSS \\
10hyq & 0.016 & SN IIP & 7.73 & $_{-0.20}^{+0.00}$ & SDSS \\
10jkx & 0.029 & SN II & 8.69 & $_{-0.04}^{+0.45}$ & SDSS \\
10kui & 0.021 & SN Ib & 9.39 & $_{-0.42}^{+0.01}$ & SDSS \\
10lnx & 0.034 & SN II & 9.45 & $_{-0.27}^{+0.00}$ & SDSS \\
10myz & 0.028 & SN IIP & 9.12 & $_{-0.08}^{+0.37}$ & SDSS \\
10npd & 0.01 & SN IIP & 6.54 & $\pm0.16$ & SDSS Phot \\
10osr & 0.024 & SN IIP & 8.04 & $\pm0.43$ & SDSS Phot \\
10qob & 0.019 & SN IIP & 8.44 & $\pm0.01$ & PanSTARRS Phot \\
10qwz & 0.02 & SN IIP & 10.29 & $\pm0.05$ & SDSS Phot \\
10rad & 0.024 & SN II & 8.90 & $\pm0.02$ & PanSTARRS Phot \\
10raj & 0.023 & SN II & 9.69 & $\pm0.02$ & PanSTARRS Phot \\
10rin & 0.032 & SN IIP & 10.13 & $\pm0.02$ & SDSS Phot \\
10rmn & 0.024 & SN IIP & 8.60 & $\pm0.01$ & SDSS Phot \\
10svt & 0.031 & SN Ic & 8.35 & $\pm0.01$ & PanSTARRS Phot \\
10tff & 0.0075 & SN IIP & 8.97 & $_{-0.22}^{+0.13}$ & SDSS \\
10tpa & 0.033 & SN IIP & 8.76 & $_{-0.08}^{+0.43}$ & SDSS \\
10ttd & 0.015 & SN IIP & 9.34 & $\pm0.06$ & SDSS Phot \\
10uiy & 0.023 & SN II & 8.64 & $_{-0.00}^{+0.15}$ & SDSS \\
10ujc & 0.032 & SN IIn & 8.44 & $\pm0.01$ & SDSS Phot \\
10vdl & 0.016 & SN IIP & 10.43 & $\pm0.03$ & PanSTARRS Phot \\
10vnv & 0.015 & SN Ib & 10.08 & $\pm0.03$ & SDSS Phot \\
10wcy & 0.024 & SN II & 10.42 & $\pm0.04$ & SDSS Phot \\
10wmf & 0.029 & SN IIP & 10.73 & $\pm0.05$ & SDSS Phot \\
10xgo & 0.034 & SN IIn & 8.22 & $\pm0.07$ & SDSS Phot \\
10xjv & 0.028 & SN IIP & 8.16 & $\pm0.09$ & SDSS Phot \\
10yna & 0.023 & SN II & 8.58 & $\pm0.03$ & SDSS Phot \\
10yow & 0.025 & SN Ic & 11.03 & $\pm0.05$ & SDSS Phot \\
10zcn & 0.02 & SN Ic & 10.18 & $\pm0.02$ & SDSS Phot \\
11bli & 0.03 & SN Ib/c & 9.37 & $_{-0.21}^{+0.14}$ & SDSS \\
11bov & 0.022 & SN Ic & 9.12 & $_{-0.15}^{+0.00}$ & SDSS \\
11cgx & 0.033 & SN II & 9.46 & $_{-0.14}^{+0.23}$ & SDSS \\
11dhf & 0.028 & SN IIb & 9.80 & $_{-0.24}^{+0.07}$ & SDSS \\
11ecp & 0.034 & SN II & 9.69 & $_{-0.21}^{+0.26}$ & SDSS \\
11eon & 0.0016 & SN IIb & 10.78 & -- & \citealt{2014MNRAS.445..899C} \\
11ftc & 0.03 & SN II & 9.23 & $\pm0.06$ & PanSTARRS Phot \\
\hline
\end{tabular}
\end{center}
\end{table*}

\begin{table*}
\caption{The host galaxy masses used to measure our rate per unit mass. We show all CCSNe, along with two SLSNe-I that were excluded by \citet{Perley2016} luminosity cut. Our masses were obtained from SDSS data products \citep[methods described in][]{2006ApJ...652...85M}, or from our own SED template fitting routines to either SDSS or PanSTARRS photometry.}
    \label{tab:host_masses_cont}
    \begin{center}
    \begin{tabular}{llllll}
        \hline
PTF Name & Redshift & SN Type & Galaxy Mass & Uncertainty & Source \\
 & & & $\mathrm{log}\, \mathrm{M}_{*} (\mathrm{M}_\odot)$ & & \\
\hline
11ftr & 0.018 & SN II & 8.01 & $\pm0.05$ & SDSS Phot \\
11fuv & 0.03 & SN IIP & 10.59 & $_{-0.15}^{+0.31}$ & SDSS \\
11gls & 0.03 & SN IIP & 10.08 & $_{-0.00}^{+0.58}$ & SDSS \\
11htj & 0.017 & SN IIP & 7.96 & $\pm0.01$ & SDSS Phot \\
11hyg & 0.03 & SN Ic & 10.82 & $\pm0.02$ & SDSS Phot \\
11iil & 0.029 & SN II & 9.60 & $\pm0.03$ & PanSTARRS Phot \\
11klg & 0.027 & SN Ic & 10.40 & $\pm0.03$ & SDSS Phot \\
11qax & 0.022 & SN IIP & 9.26 & $_{-0.02}^{+0.13}$ & SDSS \\
11qcj & 0.028 & SN Ibn & 8.58 & $_{-0.13}^{+0.03}$ & SDSS \\
12bro & 0.023 & SN IIP & 7.62 & $\pm0.15$ & SDSS Phot \\
12bvf & 0.022 & SN & 10.38 & $_{-0.44}^{+0.25}$ & SDSS \\
12cde & 0.013 & SN Ib/c & 8.02 & $\pm0.02$ & SDSS Phot \\
12cgb & 0.026 & SN II & 8.78 & $_{-0.00}^{+0.04}$ & SDSS \\
12cli & 0.024 & SN II & 10.25 & $\pm0.03$ & PanSTARRS Phot \\
12cod & 0.012 & SN II & 9.54 & $_{-0.59}^{+0.00}$ & SDSS \\
12dcp & 0.031 & SN Ic & 10.23 & $_{-0.19}^{+0.30}$ & SDSS \\
12eaw & 0.029 & SN Ib & 10.45 & $_{-0.34}^{+0.00}$ & SDSS \\
12fip & 0.034 & SN II & 8.97 & $_{-0.19}^{+0.01}$ & SDSS \\
12gfx & 0.029 & SN & 9.81 & $_{-0.00}^{+0.06}$ & SDSS \\
12gnn & 0.031 & SN II & 7.54 & $\pm0.20$ & SDSS Phot \\
12gnt & 0.029 & SN II & 9.95 & $\pm0.03$ & SDSS Phot \\
12gty & 0.176 & SLSN-I & $<9.5$ & - & SDSS Phot \\
12gzk & 0.014 & SN Ic & 6.76 & $_{-0.03}^{+0.04}$ & SDSS \\
12hap & 0.02 & SN II & 8.29 & $\pm0.36$ & SDSS Phot \\
12hni & 0.11 & SLSN-I & 8.84 & $\pm0.03$ & SDSS Phot \\ 
12hno & 0.019 & SN II & 9.63 & $\pm0.01$ & SDSS Phot \\
12hvv & 0.029 & SN Ic & 9.06 & $_{-0.15}^{+0.02}$ & SDSS \\
12iif & 0.024 & SN IIP & 10.55 & $\pm0.03$ & PanSTARRS Phot \\
12izy & 0.021 & SN II & 9.92 & $_{-0.03}^{+0.07}$ & SDSS \\
12jfp & 0.02 & SN II & 9.46 & $\pm0.05$ & SDSS Phot \\
\hline
\end{tabular}
\end{center}
\end{table*}



\bsp	
\label{lastpage}
\end{document}